\documentclass[reprint,aps,superscriptaddress,preprintnumbers,amsmath,amssymb]{revtex4-2}
\usepackage{amssymb}
\usepackage{amsmath}
\usepackage{graphicx}
\usepackage{multirow}
\usepackage{physics}
\usepackage{comment}
\usepackage{longtable}
\usepackage{color}
\usepackage{xcolor}
\usepackage{xfrac}
\usepackage{threeparttable}
\graphicspath{{./fig/}}
\usepackage[colorlinks,linkcolor=blue,anchorcolor=black,citecolor=blue,urlcolor=black]{hyperref}

\newcommand{\br}{\mathbf{r}}
\newcommand{\brp}{\mathbf{r}^\prime}

\begin{document}
\newcommand{\tny}[1]{\mbox{\tiny $#1$}}
\newcommand{\eqn}[1]{\mbox{Eq.\hspace{1pt}(\ref{#1})}}
\newcommand{\eqs}[2]{\mbox{Eq.\hspace{1pt}(\ref{#1}--\ref{#2})}}
\newcommand{\eqsu}[2]{\mbox{Eqs.\hspace{1pt}(\ref{#1},\ref{#2})}}
\newcommand{\eqtn}[2]{\begin{equation} \label{#1} #2 \end{equation}}
\newcommand{\func}[1]{#1 \left[ \rho \right] }
\newcommand{\mfunc}[2]{#1_{#2} \left[ \rho \right] }
\newcommand{\mmfunc}[3]{#1_{#2} \left[ #3 \right] }
\newcommand{\mmmfunc}[4]{#1_{#2}^{#3} \left[ #4 \right] }
\newcommand{\pot}[1]{v_{\rm #1}}
\newcommand{\spot}[2]{v_{\rm #1}^{#2}}
\newcommand{\code}[1]{\texttt{#1}}
\newcommand{\beq}{\begin{equation}}
\newcommand{\eeq}{\end{equation}}
\newcommand{\bea}{\begin{eqnarray}}
\newcommand{\eea}{\end{eqnarray}}

\def\brpppp{{\mathbf{r}^{\prime\prime\prime\prime}}}
\def\brppp{{\mathbf{r}^{\prime\prime\prime}}}
\def\brpp{{\mathbf{r}^{\prime\prime}}}
\def\brp{{\mathbf{r}^{\prime}}}
\def\bzp{{\mathbf{z}^{\prime}}}
\def\bxp{{\mathbf{x}^{\prime}}}
\def\tp{{{t}^{\prime}}}
\def\tpp{{{t}^{\prime\prime}}}
\def\tppp{{{t}^{\prime\prime\prime}}}

\def\tbr{{\tilde{\mathbf{r}}}}
\def\bk{{\mathbf{k}}}
\def\br{{\mathbf{r}}}
\def\bz{{\mathbf{z}}}
\def\bx{{\mathbf{x}}}
\def\bR{{\mathbf{R}}}
\def\bM{{\mathbf{M}}}
\def\bP{{\mathbf{P}}}
\def\bT{{\mathbf{T}}}
\def\bK{{\mathbf{K}}}
\def\bA{{\mathbf{A}}}
\def\bB{{\mathbf{B}}}
\def\bX{{\mathbf{X}}}
\def\bY{{\mathbf{Y}}}
\def\bP{{\mathbf{P}}}
\def\bI{{\mathbf{I}}}
\def\d{{\mathrm{d}}}
\def\rhor{{\rho({\bf r})}}
\def\rhorp{{\rho({\bf r}^{\prime})}}
\def\rhoi{{\rho_I}}
\def\rhoii{{\rho_{II}}}
\def\rhoj{{\rho_J}}
\def\rhoir{{\rho_I({\bf r})}}
\def\rhoiir{{\rho_{II}({\bf r})}}
\def\rhojr{{\rho_J({\bf r})}}
\def\rhoirp{{\rho_I({\bf r}^{\prime})}}
\def\rhojrp{{\rho_J({\bf r}^{\prime})}}
\def\sumi{{\sum_I^{N_S}}}
\def\sumj{{\sum_J^{N_S}}}
\def\im{{\operatorname{Im}}}

\def\etal{{\it et al.}}
\def\vdw{{van der Waals}}
\def\vw{{von Weizs\"{a}cker}}
\def\qe{{\sc Quantum ESPRESSO}}
\def\se{{Schr\"{o}dinger equation}}
\def\ses{{Schr\"{o}dinger equations}}
\def\bnabla{{\boldsymbol{\nabla}}}
\def\bchi{{\boldsymbol\chi}}
\def\bLambda{{\boldsymbol\Lambda}}
\def\bDelta{{\boldsymbol\Delta}}

\title{Nonlocal free-energy density functional for warm dense matter}
\author{Cheng Ma}
\affiliation{Key Laboratory of Material Simulation Methods $\&$ Software of Ministry of Education, College of Physics, Jilin University, Changchun 130012, China. }
\affiliation{State Key Lab of Superhard Materials, College of Physics, Jilin University, Changchun 130012, China. }
\author{Min Chen}
\affiliation{Key Laboratory of Material Simulation Methods $\&$ Software of Ministry of Education, College of Physics, Jilin University, Changchun 130012, China. }

\author{Yu Xie}
\affiliation{Key Laboratory of Material Simulation Methods $\&$ Software of Ministry of Education, College of Physics, Jilin University, Changchun 130012, China. }

\author{Qiang Xu}
\email{xq@calypso.cn}
\affiliation{Key Laboratory of Material Simulation Methods $\&$ Software of Ministry of Education, College of Physics, Jilin University, Changchun 130012, China. }
\author{Wenhui Mi}
\email{mwh@jlu.edu.cn}
\affiliation{Key Laboratory of Material Simulation Methods $\&$ Software of Ministry of Education, College of Physics, Jilin University, Changchun 130012, China. }
\affiliation{State Key Lab of Superhard Materials, College of Physics, Jilin University, Changchun 130012, China. }
\affiliation{International Center of Future Science, Jilin University, Changchun 130012, China.}
\author{Yanchao Wang}
\email{wyc@calypso.cn}
\affiliation{Key Laboratory of Material Simulation Methods $\&$ Software of Ministry of Education, College of Physics, Jilin University, Changchun 130012, China. }
\affiliation{State Key Lab of Superhard Materials, College of Physics, Jilin University, Changchun 130012, China. }
\author{Yanming Ma}
\email{mym@jlu.edu.cn}
\affiliation{Key Laboratory of Material Simulation Methods $\&$ Software of Ministry of Education, College of Physics, Jilin University, Changchun 130012, China. }
\affiliation{State Key Lab of Superhard Materials, College of Physics, Jilin University, Changchun 130012, China. }
\affiliation{International Center of Future Science, Jilin University, Changchun 130012, China.}

\begin{abstract}
Finite-temperature orbital-free density functional theory (FT-OFDFT) holds significant promise for simulating warm dense matter due to its favorable scaling with both system size and temperature. However, the lack of the numerically accurate and transferable noninteracting free energy functionals results in a limit on the application of FT-OFDFT for warm dense matter simulations. Here, a nonlocal free energy functional, named XWMF, was derived by line integrals for FT-OFDFT simulations. Particularly, a designed integral path, wherein the electronic density varies from uniform to inhomogeneous, was employed to accurately describe deviations in response behavior from the uniform electron gas. The XWMF has been benchmarked by a range of warm dense matter systems including the Si, Al, H, He, and H-He mixture. The simulated results demonstrate that FT-OFDFT within XWMF achieves remarkable performance for accuracy and numerical stability. It is worth noting that XWMF exhibits a low computational cost for large-scale $ab~initio$ simulations, offering exciting opportunities for the realistic simulations of warm dense matter systems covering a broad range of temperatures and pressures.
\end{abstract}
\maketitle

\section{introduction}
As a bridge between cold condensed matter and hot plasma, warm dense matter (WDM) represents an extreme state comprising a matter regime characterized by high temperatures ($\sim$10 eV) and high pressures ($\sim$1 Mbar or higher). The WDM has attracted tremendous attention in various fields, including planet science, high-energy-density physics, and materials science, due to its ubiquity throughout nature, such as exo-planet interiors, the path to inertial confinement fusion, and neutron stellar atmospheres \cite{graziani2014frontiers,david2021warm}. In WDM regime, electrons are usually in a highly excited state and partially degenerate, exhibiting non-negligible quantum effects. Therefore, $ab~initio$-based simulations have become an indispensable approach for describing the phenomena and mechanisms of WDM. 

Currently, several approaches including finite-temperature Kohn-Sham density functional theory (FT-KSDFT) \cite{hk1964,kohn1965self,mermin1965thermal,parr1989density,mazevet2007ab,holst2008thermophysical,wang2013wide,kang2018dynamic}, path-integral Monte Carlo (PIMC) \cite{pollock1984,ceperley1995path,driver2012all,hu2011first}, extended first-principles molecular dynamics (ext-FPMD) \cite{zhang2016extended,blanchet2020requirements}, and stochastic density functional theory (sDFT) \cite{baer2013self,white2020fast,cytter2018stochastic,liu2022plane} have been established and yielded successes to simulate WDM. Particularly, a combination of FT-KSDFT and molecular dynamics (MD) has been proposed as an ideal framework for simulations of WDM, where electrons and ions are treated by quantum mechanical KSDFT and classical MD, respectively. However, the heavy computational demands for FT-KSDFT simulations at high temperature make its use for WDM problematic \cite{cytter2018stochastic}. Even with modern supercomputers FT-KSDFT can only used to treat the WDM at relatively low temperature without further tricks.

The finite-temperature orbital-free density functional theory (FT-OFDFT) \cite{wang2000orbital,karasiev2012issues,Karasiev2012GGA,witt2018orbital,mi2023} combined with MD offers a promising alternative to simulate the WDM because its computational cost grows linearly with system size and almost irrespective of temperatures. It is necessary to draw attention that the performance of FT-OFDFT critically depends on the noninteracting free energy density functional (FEDF). In this context, the development of accurate and reliable FEDF is highly desirable.

The early FEDFs based on finite temperature Thomas-Fermi (TF) theory \cite{Fermi1927,Thomas1927,feynman1949equations,bartel1985263extended,perrot1979gradient} were derived and 
successfully applied to warm dense plasma at ultra-high temperatures \cite{lambert2006very,lambert2006structural,danel2006equation}. Later, several semilocal FEDFs such as VT84F \cite{Karasiev2013VT84F} and LKTF \cite{luo2020towards} have been derived from generalized-gradient approximation (GGA) \cite{Karasiev2012GGA}. These GGA-FEDFs show significant accuracy improvement in WDM simulations, compared to TF-FEDFs \cite{zhang2022first,kang2020two,moldabekov2021relevance}. While it is the fact that the GGA functionals are lack of nonlocality and cannot reproduce the exact Lindhard response \cite{Lindhard1954,sjostrom2013SD13,moldabekov2023imposing, mi2023}. A recent nonlocal FEDF, denoted as WTF \cite{sjostrom2013SD13}, has been constructed by a direct extension of the Wang-Teter kinetic energy density functional \cite{wang1992kinetic} and shown high accuracy across a broad range of temperatures \cite{giuliani2008quantum,wang2000orbital}. However, the WTF was found to be numerically problematic under specific conditions \cite{sjostrom2014SD14}. To address this issue, a modified functional, WTF$\beta$vW \cite{sjostrom2014SD14}, was later proposed and exhibited remarkable performance in WDM simulations. Despite the advances in the FEDF within FT-OFDFTs over past decades, there is huge room for future development, and any progress could impact the FT-OFDFT application in WDM simulations.

In this work, we derived an advanced nonlocal FEDF using line integrals, following the theoretical formalism of Xu-Wang-Ma (XWM) kinetic energy density functional \cite{xu2019nonlocal}. Numerical assessments reveal that XWM-FEDF (XWMF) exhibits numerical stability and high accuracy, as demonstrated by fact that results calculated by XWMF were in good agreement with FT-KSDFT results.

The reminder of this manuscript is organized as follows: Sec. \ref{sec:method} offers the definition and construction of XWMF functional. The computational details are provided in Sec. \ref{sec:details}. The calculated results and discussion are presented in Sec. \ref{sec:disscuss}, followed by conclusions in Sec. \ref{sec:conclusions}.

\section{Methods}
\label{sec:method}
In contrast to the simulation at zero temperature, the grand canonical potential instead of energy \cite{mermin1965thermal,parr1989density} is the critical physical quantity of interest at finite temperature. The grand potential ($\Omega$) of elections can be expressed as a density functional:
\beq
\Omega[\rho;T]=F[\rho;T]+\int d^3\br v_{ext}(\br)\rho(\br)-\mu N_e,\label{eq:1}
\eeq
where $\rho$, $T$, $v_{ext}$, $\mu$, and $N_e$ are the electron density, absolute temperature, external potential, chemical potential, and number of electrons, respectively. The universal functional $F[\rho]$ is an unknown term but could be decomposed into three terms \cite{parr1989density}:
\beq
F[\rho;T] = F_s[\rho;T]+F_{H}[\rho]+F_{XC}[\rho;T],\label{eq:2}
\eeq
where $F_s$, $F_{H}$, and $F_{XC}$ denote the noninteracting free energy, the classical Coulomb energy (Hartree energy), and exchange-correlation energy, respectively. Generally, $F_{H}$ explicitly depends on the electron density and can be calculated by {$F_{H}[\rho] = \int\int \frac{\rho(\br)\rho(\br')}{|\br-\br'| }d^3\br d^3\br'$. While the approximate functional forms should be required to express the $F_{s}[\rho;T]$ and $F_{XC}[\rho;T]$. Several approximate forms of $F_{XC}$, including local-density approximation \cite{karasiev2014accurate}, GGA \cite{karasiev2018nonempirical}, hybrids \cite{mihaylov2020thermal}, and meta-GGA \cite{karasiev2022meta} functionals have been proposed for finite-temperature simulations. Moreover, the adiabatic approximation ($F_{XC}[\rho;T] \approx E_{XC}[\rho]$) has also been widely used in such simulations \cite{moldabekov2021relevance,moldabekov2022benchmarking,moldabekov2023assessing}. These exchange-correlation functionals usually yield a reasonable results since the $F_{XC}$ is an order of magnitude smaller than $F$ \cite{Karasiev2012GGA}. 

Since the $F_s$ has the same order of magnitude as the functional $F$, it plays a crucial role in determining accuracy of FT-OFDFT simulations \cite{Karasiev2012GGA}. Generally, the nonlocal FEDF can be written as \cite{wang1992kinetic,sjostrom2013SD13,sjostrom2014SD14}:
\beq
F_s[\rho;T] = F_s^{TF}[\rho;T]+F_s^{vW}[\rho;T]+F_s^{NL}[\rho;T],\label{eq:3}
\eeq
where $F_s^{TF}$ is the TF-FEDF, which can be derived from the finite-temperature TF theory: 
\beq
F_s^{TF}[\rho;T] = \int f_s^{TF}(\br;T)d^3\br,\label{eq:4}
\eeq
where $f_s^{TF}(\br;T)=t_0^{TF}(\br)k(\tau,\br)$ is the TF free energy density. $t_0^{TF}=\frac{3}{10}(3\pi^2)^{2/3}\rho^{5/3}(\br)$ is the TF kinetic energy density at 0K, and the factor $k(\tau)$ has an analytical form \cite{perrot1979gradient,valentin2015improved} at finite temperature. $\tau = \frac{2k_BT}{[3\pi^2 \rho]^{2/3} }$ and $k_B$ denote the reduced temperature and Boltzmann constant, respectively.

The second term in Eq.~(\ref{eq:3}) is von Weizs\"acker (vW) functional \cite{weizsacker1935}. At finite temperature, three vW-functional forms have been proposed: (i) the adiabatic approximation form adopted as vW kinetic energy \cite{weizsacker1935} $ F_s^{vW}[\rho;T] = T_s^{vW}[\rho]\equiv\int \frac{|\nabla\rho(\br)|^2}{8\rho(\br)} d^3\br$; (ii) The vW functional scaled by a  reduced temperature function derived from finite-temperature gradient correction \cite{perrot1979gradient,bartel1985263extended,Karasiev2012GGA}; and (iii) $\beta$vW form employed in WTF$\beta$vW \cite{sjostrom2014SD14}. Note that the adiabatic vW functional ($F_s^{vW}=T_s^{vW}$) is adopted in this work because of its correct asymptotic behavior in the linear response of the uniform electron gas \cite{sjostrom2013SD13}.

The third term in Eq.~(\ref{eq:3}) is the nonlocal part free energy (NLFE). Following the formalism of XWM kinetic energy density functional \cite{xu2019nonlocal}, NLFE can be derived from the line integral:
\bea
F_s^{NL}[\rho;T]=&&F_s^{NL}[\rho_{t=0};T]\nonumber\\
+&&\int d^3\br\int_0^1 dt \frac{\delta F_s^{NL}[\rho;T]}{\delta \rho_t(\br)}\frac{d \rho_t(\br)}{dt}.\label{eq:5}
\eea
Note that the first-order functional derivative in the integrand can also be evaluated from the line integral: 
\bea
\frac{\delta F_s^{NL}[\rho;T]}{\delta \rho(\br)} =&& \frac{\delta F_s^{NL}[\rho;T]}{\delta \rho(\br)}|_{\rho_{t'=0}} \nonumber\\
+&& \int d^3\br'\int_0^1 dt' \frac{\delta^2 F_s^{NL}[\rho;T]}{{\delta\rho(\br)\delta\rho_{t'}(\br')}} \frac{d \rho_{t'}(\br)}{dt'}.\label{eq:6}
\eea

In principle, one can derive an exact NLFE functional expression according to Eqs~(\ref{eq:5}) and (\ref{eq:6}) if the second derivative of NLFE, denoted as $(F_s^{NL})''(\br,\br')\equiv\frac{\delta^2F_s^{NL}}{\delta\rho(\br)\delta\rho(\br')}$, is known along a given density path $\rho_t$. However, the exact condition is only available by the Lindhard theory \cite{Lindhard1954,giuliani2008quantum} at uniform electron gas, where $\rho = \rho_0$. Taking this into consideration, the density integral path is selected by the following form \cite{xu2019nonlocal,chai2007orbital}:
\beq
\rho_t(\br)= \rho_0+t[\rho(\br)-\rho_0],\label{eq:7}
\eeq
where $t\in[0,1]$. The peculiar path in Eq.~(\ref{eq:7}) has proven to be successfully applied in describing near-free electron gas systems \cite{xu2019nonlocal}. Furthermore, the initial values in Eqs.~(\ref{eq:5}) and (\ref{eq:6}), $F_s^{NL}[\rho_{t=0}]$ and $\frac{\delta F_s^{NL}}{\delta \rho(\br)}|_{\rho_{t'=0}}$, naturally become zero for this density path. However, the general form of the second order derivative of NLFE in Eq.~(\ref{eq:6}) is unknown along this path except for the uniform electron gas \cite{Lindhard1954}:
\bea 
(F_s^{NL})''[\rho_0;T](\br,\br') =&&-\chi_{NL}^{-1}[\rho_0;T](\br,\br') \nonumber \\
 \equiv&& \rho_0^{-1/3}G(k_F^0|\br-\br'|;T),\label{eq:8} 
\eea
where ${\chi}^{-1}_{NL}=\chi_{L}^{-1}-{\chi}^{-1}_{TF}-{\chi}^{-1}_{vW}$. $\chi_{L}^{-1}$, ${\chi}^{-1}_{TF}$, and ${\chi}^{-1}_{vW}$ represent the Lindhard-reponse, TF-reponse, and vW-reponse functions, respectively \cite{Lindhard1954}. The $G$-function term of Eq.~(\ref{eq:8}) is a dimensionless function and has an analytical form in momentum space, see supplemental material \footnotemark[111]. For a general inhomogeneous electron density, we extended Eq~(\ref{eq:7}) to a general form by introducing $\rho_0^{-1/3}\to\rho^{-1/6}(\br)\rho^{-1/6}(\br')$ and $k_F^0\to\varepsilon_\gamma(\br,\br')=[\frac{k_F^\gamma(\br)+k_F^\gamma(\br')}{2}]^{1/\gamma}$:
\bea
(F_s^{NL}&&)''[\rho;T](\br,\br')= \nonumber \\ &&I^{\kappa}(\br)\rho^{-\frac{1}{6}}(\br)G(\varepsilon_\gamma|\br-\br'|;T)\rho^{-\frac{1}{6}}(\br')I^{\kappa}(\br'),\label{eq:10}
\eea
where $I^\kappa(\br)=[\rho(\br)/\rho_0]^\kappa$ and $\kappa$ is an adjustable parameter. Note that Eq.~(\ref{eq:10}) will naturally degenerate into Eq.~(\ref{eq:8}) when the electron density is a uniform electron gas [$\rho_{t=0}=\rho_0$]. Therefore, the additional term $I^\kappa(\br)I^\kappa(\br')$ in Eq.~(\ref{eq:10}) is introduced to tune the second derivative of NLFE for inhomogeneous electron gas along the integral path ($t\neq 0$). To reduce the complexity, we simplify the density-dependent $G$-function in Eq.~(\ref{eq:10}) by a Taylor series expansion:
\bea
G(\varepsilon_\gamma|\br-\br'|;T) \simeq && G(k_F^*|\br-\br'|;T) \nonumber\ \\ 
&&+\frac{\partial G(\varepsilon_\gamma|\br-\br'|;T)}{\partial \rho(\br)}\mid_{\rho^*}\Delta\rho(\br)\nonumber \\
&&+ \frac{\partial G(\varepsilon_\gamma|\br-\br'|;T)}{\partial \rho(\br')}\mid_{\rho^*}\Delta\rho(\br'),\nonumber\\
\label{eq:11}
\eea
where $\rho^*$ is a reference charge density, where the Taylor expansions around, and $\Delta\rho(\br)=\rho(\br)-\rho^*$. So far, the nonlocal part of kinetic functional can be derived by combining Eqs.~(\ref{eq:5}-\ref{eq:11}) :
\beq
F_s^{NL}[\rho;T] = F^{NL}_0[\rho;T] + F^{NL}_1[\rho;T],\label{eq:xwmfx}
\eeq
where the zero-order NLFE ($F_0^{NL}$) is given by:
\beq
F_0^{NL}[\rho;T] = \int\int d^3\br d^3\br' \rho^{\kappa+\frac{5}{6}}(\br)\omega_0(\br,\br')\rho^{\kappa+\frac{5}{6}}(\br'),\label{xwmf0}
\eeq
where $\omega_0=\frac{18G(k_F^*|\br-\br'|;T)}{(6\kappa+5)^2\rho_0^{2\kappa}}$. Note that $F_0^{NL}$ shows the same form as nonlocal part of WTF when $\kappa=0$ and $\rho^*=\rho_0$. The first-order term is given by: 
\bea
F_1^{NL}[\rho;T]=&&\int\int d^3\br d^3\br' \rho^{\kappa+\frac{11}{6}}(\br)\omega_{11}(\br,\br')\rho^{\kappa+\frac{5}{6}}(\br')\nonumber\\
+&&\int\int d^3\br d^3\br' \rho^{\kappa+\frac{5}{6}}(\br)\omega_{12}(\br,\br')\rho^{\kappa+\frac{5}{6}}(\br'),\nonumber\\\label{xwmf1}
\eea
where $\omega_{11}=\frac{G'(\br,\br';T)}{(\kappa+{5/6})(\kappa+{11/6})\rho_0^{2\kappa}}$, $\omega_{12}=-\frac{\rho^*G'(\br,\br';T)}{(\kappa+{5/6})^2\rho_0^{2\kappa}}$, and $G'\equiv\frac{\partial G(\varepsilon_\gamma|\br-\br'|;T)}{\partial \rho(\br)}\mid_{\rho^*}$. In practice, a compact form of XWMF is used for implementation with high computational efficiency by combining Eq.~(\ref{xwmf0}) and the second line of Eq.~(\ref{xwmf1}) \cite{xu2022nonlocal}:
\bea
F_s^{NL}[\rho;T]=&&\int\int d^3\br d^3\br' \rho^{\kappa+\frac{5}{6}}(\br)\omega_{1}(\br,\br')\rho^{\kappa+\frac{5}{6}}(\br')\nonumber\\
+&&\int\int d^3\br d^3\br' \rho^{\kappa+\frac{11}{6}}(\br)\omega_{2}(\br,\br')\rho^{\kappa+\frac{5}{6}}(\br'),\nonumber\\\label{xwmf}
\eea
where $\omega_1=\omega_0+\omega_{12}$, and $\omega_2=\omega_{11}$. We should point out here that XWMF will naturally degenerate into XWM kinetic energy density functional \cite{xu2019nonlocal} at $T=0$~K. The detailed derivation and implementation of XWMF are provided in the supplemental material \footnotemark[111].

\footnotetext[111]{See Supplemental Material at [URL will be inserted by publisher]}

\section{Computational details}
\label{sec:details}
The various FEDFs including TF, WTF, WTF$\beta$vW, and XWMF have been implemented in ATLAS \cite{mi2016atlas,shao2018large}. To validate the accuracy of XWMF functional, we conducted static lattice calculations and $ab~initio$ molecular dynamics (AIMD) using both FT-OFDFT and FT-KSDFT, employing identical pseudopotentials. For uniformity, we employed the adiabatic approximation using the adiabatic Perdew-Zunger local density approximation \cite{Perdew1981PZ} as the exchange-correlation (XC) functional across all calculations. For XWMF, the parameters of $\rho^*=\rho_0$ and $\kappa=0$ are adopted for all cases. In all FT-KSDFT calculations, appropriate band numbers were chosen to encompass all bands with occupations exceeding $5\times10^{-5}$, as detailed in the supplemental material \footnotemark[111].

For the static lattice equation of state (EOS) calculations, BLPS local pseudopotentials \cite{Zhou2004BLPS1,Huang2008BLPS2} were adopted for face-centered cubic (FCC) Al and cubic diamond (CD) Si; Heine-Abarenkov pseudopotentials (HAPPs) \cite{heine1964new,goodwin1990pseudopotential,Karasiev2012GGA} was employed for face-centered cubic (FCC) H. The grid spacing of 0.08,  0.08, and 0.05 \AA~for Al, Si, and H was employed in ATLAS to ensure convergence of the free energy to within 1~meV/atom. FT-KSDFT calculations were performed using CASTEP 8.0 \cite{CASTEP2005}, employing a kinetic energy cutoff of 1000~eV for Al and Si, and 4000~eV for H. The k-point meshes were generated via the Monkhorst-Pack method \cite{Monkhorst1976} with a spacing of 0.016 \AA$^{-1}$.

The AIMD calculations were performed to obtain the EOS of hydrogen (H), Helium (He), and the H-He mixture. The HAPPs \cite{heine1964new,goodwin1990pseudopotential,Karasiev2012GGA} with $r_{cut}=0.25$ Bohr were adopted for H with a bulk density of less than 8.0 $g/cm^3$. To simulate denser H and He, the Troullier-Martins norm-conserving nonlocal pseudopotentials (NLPPs) \cite{Troullier1991} were generated by FHI98PP \cite{fuchs1999fhi98} containing only $s-$channel with 0.10 and 0.25 Bohr cutoffs. In FT-OFDFT calculations, 0.05 and 0.03 \AA~grid spacings were employed for HAPPs and NLPPs. The FT-KSDFT-AIMD was performed with ABACUS 3.5.3 \cite{chen2010abacus,li2016large}. The kinetic energy cutoff and k-point meshes were varied from 200 to 300 Ry and $\Gamma$-only to $3\times3\times3$. The number of atoms was 108 for H less than 8.0 $g/cm^3$, and 125 atoms for the denser case. The time step varied from 0.005 to 0.1 fs. All AIMD were performed with the NVT ensemble controlled by Nose–Hoover thermostat \cite{nose1984a,hoover1985canonical}. Each system was run of 10000 steps for initial equilibration; after that, pressures were averaged over the next 10000 steps.

\section{Results and discussion}
\label{sec:disscuss}

\subsection{Static lattice equations of state}
\begin{figure}[htbp]
\includegraphics[width=0.95 \linewidth]{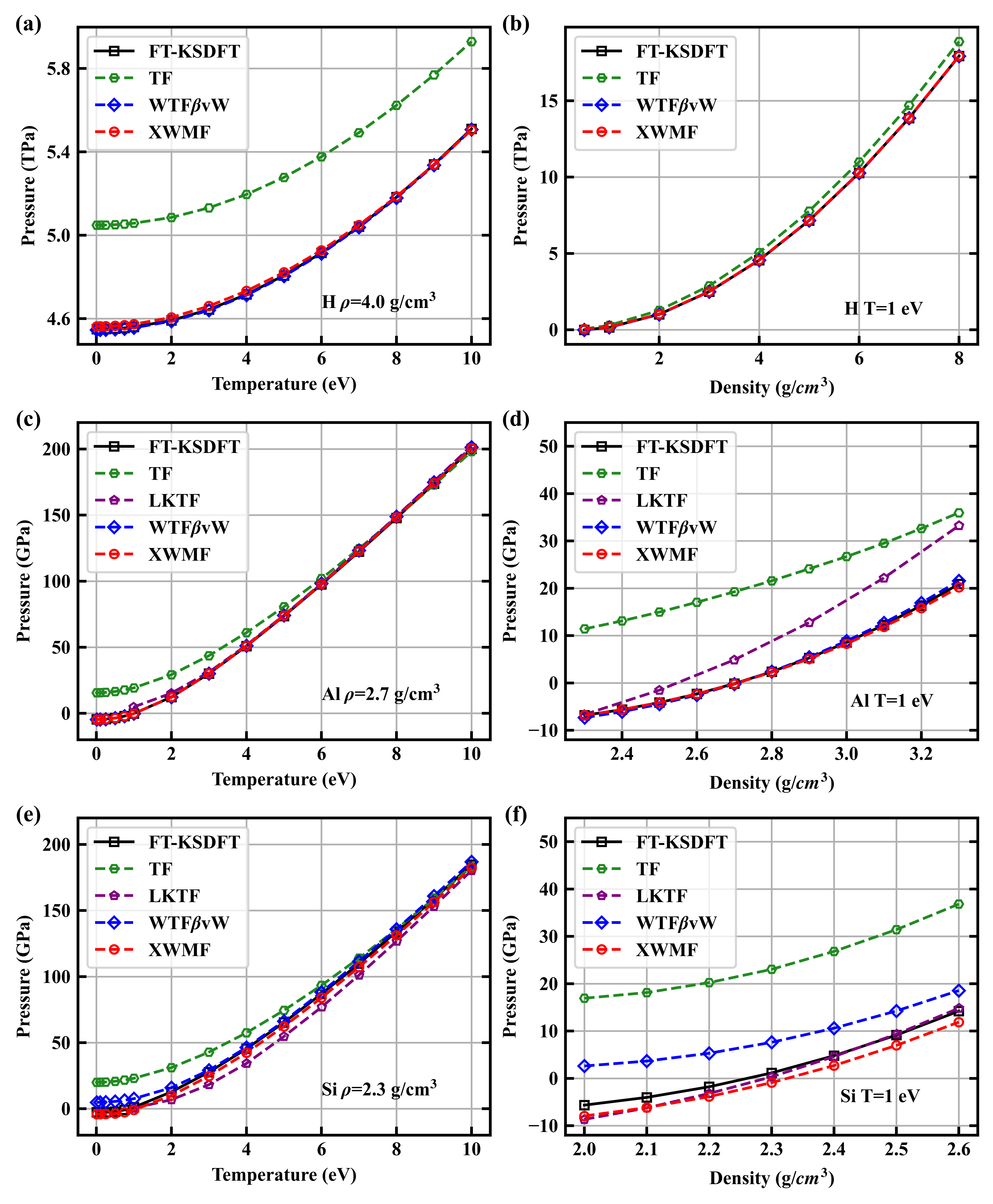}
\caption{\label{fig:Al_12} Electronic pressure prediction comparison for various free-energy functionals compared with FT-KSDFT results for the static lattice of  (a-b) FCC H, (c-d) FCC Al, and (e-f) CD Si. The LKTF data are adopted from Ref.\cite{luo2020towards}. }
\end{figure}
\label{IVA}

To evaluate the accuracy of XWMF, we initially applied it to obtain the static lattice equations of state (EOS) of FCC H, FCC Al, and CD Si. Comparisons of the curves of pressure versus temperature and density obtained by FT-OFDFT using XWMF, TF, LKTF, and WTF$\beta$vW, along with the FT-KSDFT results were shown in Fig.~\ref{fig:Al_12}. Overall, the nonlocal functionals were superior to the local (TF) and semilocal (LKTF) functionals for all the considered systems. It can be clearly seen that the curves obtained by FT-KSDFT were well reproduced by the nonlocal functionals of WTF$\beta$vW and XWMF across the considered temperature and density ranges for Al and H with FCC structures, as presented in Fig.~\ref{fig:Al_12} (a)--(d). For CD Si, the pressure deviations evaluated by FT-OFDFT within XWMF with respect to those of FT-KSDFT were generally small for the considered temperatures and densities. This was in contrast to WTF$\beta$vW which produces a large deviation at low temperatures ranging from 0.01 to 1 eV, as shown in Fig.~\ref{fig:Al_12} (e)--(f). For example, the pressure deviation between WTF$\beta$vW and FT-KSDFT is 6.47 GPa at $\rho=2.3 g/cm^3$ and $T=1$~eV, while XWMF has contributed to a significant reduction of this deviation to 2.05~GPa. These results illustrated that FT-OFDFT within XWMF can give credibility for simulations of WDM over a wide range of pressures and temperatures.

\subsection{Ab initio molecular dynamics}
The FT-OFDFT has been proved to yield a powerful predictive capability in determining the thermophysical properties, offering the same accuracy as the FT-KSDFT. Thus, 
the XWMF makes FT-OFDFT to be sufficient in both accuracy and cost savings for the simulations of WDM, allowing us to investigate the properties of warm dense H and H-He mixtures under relevant planetary conditions.

\begin{table*}[htbp]
\caption{\label{tab:HMD} The evaluated pressure (in Mbar) of warm dense H with various densities (in $g/cm^3$) via FT-KSDFT-MD, and FT-OFDFT-MD simulations using WTF/WTF$\beta$vW/XWMF  at $T$ = 25 and 50 kK. $^a$The FT-KSDFT and LKTF results are adopted from Ref.\cite{luo2020towards} .}
\begin{ruledtabular}
\begin{tabular}{rllllllllll}
       &  Temperature & Density  &  FT-KSDFT & FT-KSDFT$^a$  & LKTF$^a$ & WTF     & WTF$\beta$vW & XWMF   \\ 
\hline 
       &              &  0.6     &   2.1  & 2.1         &  1.9     &   -     &  -            &   1.9 \\
       &              &  1.0     &   4.9  & 5.0         &  4.6     &   -     &  5.1          &   4.8 \\
       &     25kK     &  2.0     &  16.7  & 16.9        &  16.3    &   -     &  16.9         &  16.7 \\
       &              &  4.0     &  58.8  & 59.1        &  58.5    &  59.3   &  59.0         &  58.7 \\
       &              &  8.0     & 205.7  & 207.2       &  205.8   &  206.2  &  205.8        & 205.7 \\
\hline       
       &              &  0.6     &   3.9  & 3.9         &  3.6     &   -     &  4.1          &   3.6 \\
       &              &  1.0     &   7.9  & 8.0         &  7.5     &   -     &  8.1          &   7.6 \\
       &     50kK     &  2.0     &  22.5  & 22.7        &  22.2    &   -     &  22.9         &  22.4 \\
       &              &  4.0     &  70.1  & 70.6        &  69.9    &  71.1   &  70.5         &  70.1 \\
       &              &  8.0     &  228.0 & 229.5       &  228.3   &  229.0  &  228.5        & 228.0 \\
\end{tabular}
\end{ruledtabular}
\end{table*}
\begin{figure}[htp]
    \includegraphics[width=0.85 \linewidth]{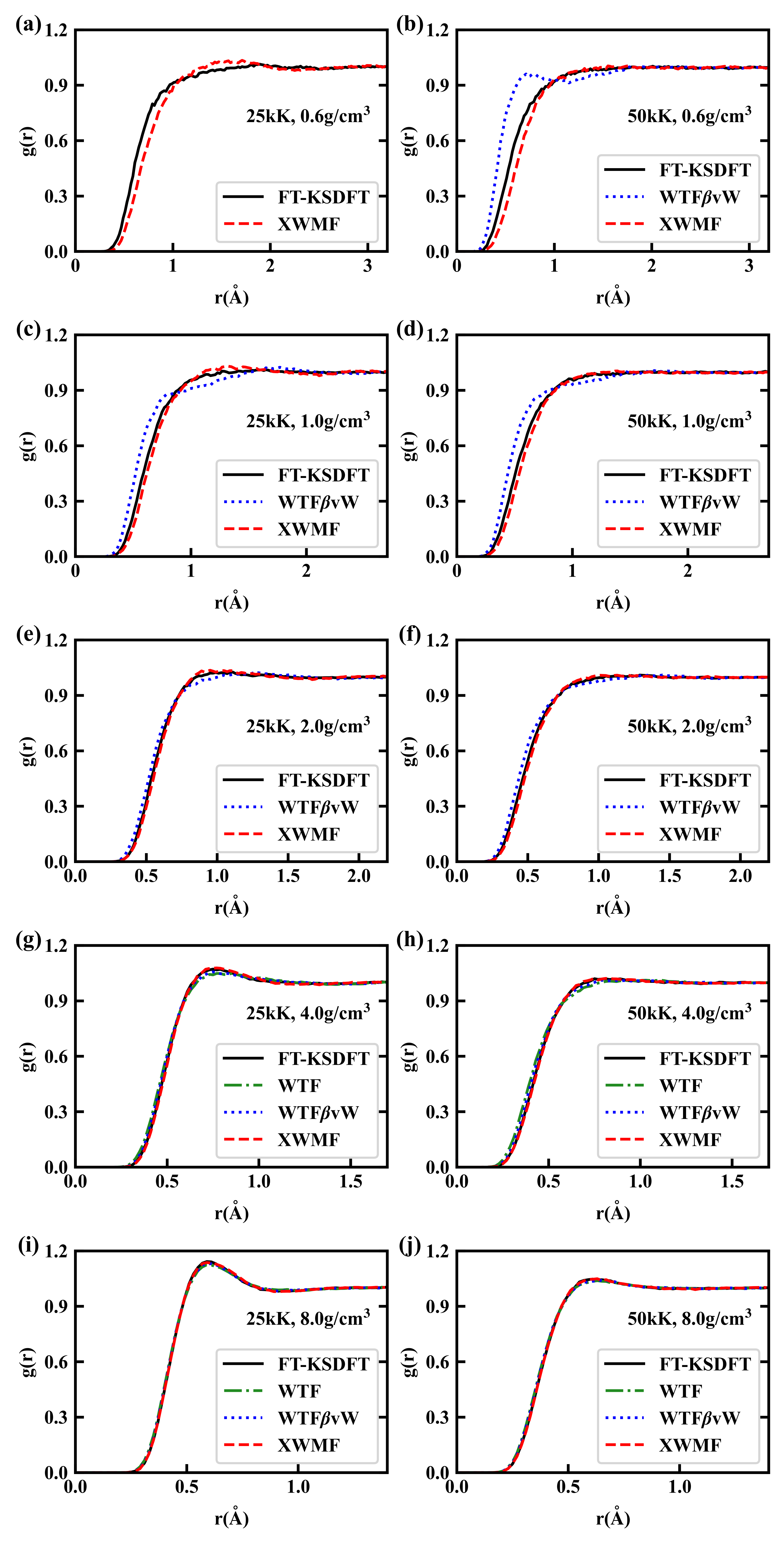}
    \caption{\label{fig:rdf}The pair distribution functions of H calculated by FT-KSDFT, and FT-OFDFT within WTF, WTF$\beta$vW, and XWMF at 25 kK (left) and 50 kK (right). } 
\end{figure}

To illustrate the validity of XWMF, the equilibrium pressures of H at 25 and 50 kK were calculated by FT-KSDFT-MD and FT-OFDFT-MD using WTF, WTF$\beta$vW, and XWMF. The calculated results, along with those obtained by LKTF and FT-KSDFT in Ref.\cite{luo2020towards} were presented in Table~\ref{tab:HMD}. Generally, the accuracy of XWMF implemented in FT-OFDFT-MD achieves comparable or better than previous FEDFs. In particular, there was remarkable agreement between the XWMF and WTF$\beta$vW for the warm dense H at 2.0 $g/cm^3$ and 25kK. The estimated equilibrium pressures were 16.7 and 16.9 Mbar for XWMF and WTF$\beta$vW, respectively. The results were well consistent with that obtained by FT-KSDFT, but LKTF given a relatively large discrepancy with respect to FT-KSDFT (0.4 Mbar).

Generally, the pair distribution functions are of importance to represent the structure-properties of matter, such as local bonding and composition. Therefore, we calculated the pair distribution functions for warm condense H with various densities at 25 and 50 kK. Just as shown in Fig.~\ref{fig:rdf}, our estimate of the pair distribution functions using FT-OFDFT-MD with all FEDFs were in perfect agreement with the estimates of FT-KSDFT-MD for the WDM of H at high densities. However, the stark differences between XWMF and other previous FEDFs were observed at lower densities. XWMF provides better predictions than WTF$\beta$vW for pair distribution functions at the densities of 0.6--2.0 $g/cm^3$. Our findings indicated that the XWMF has a dramatic improvement in numerical accuracy for predicting equilibrium pressures and pair distribution functions of WDM. 

Besides accuracy, the numerical stability is an alternative important issue for the FEDFs. According to our results, FT-OFDFT-MD simulations using the WTF and WTF$\beta$vW FEDFs exhibit numerical instabilities. The situation was more dramatic for the cases of WDM at lower densities. While XWMF enabled to effectively overcome this problem to a large extent, as evidenced by fact that the simulations for all the considered systems with various densities were successfully converged.

\begin{figure}[htp]
\includegraphics[width=0.8 \linewidth]{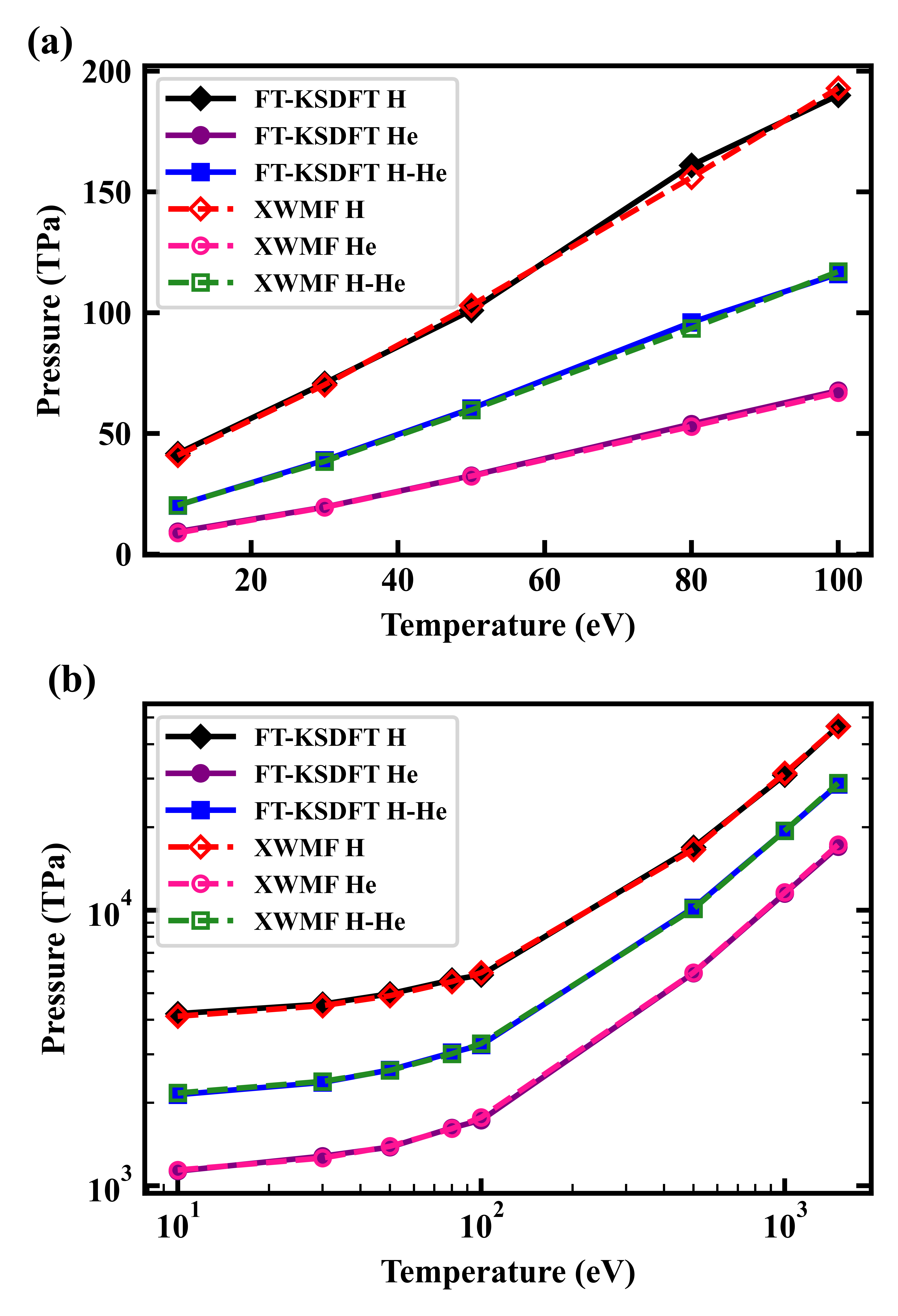}
\caption{\label{fig:MD3} Pressures versus temperatures for H, He, and H-He mixture with 0.4 mass abundance of H calculated by FT-OFDFT using XWMF and FT-KSDFT at (a) $\rho = 10~g/cm^3$ and (b) $\rho = 160~g/cm^3$. FT-KSDFT results are adopted from Ref. \cite{dai2010quantum}.}
\end{figure}

It is well-known that the H and He are ubiquitous in giant planets and numerous exoplanets 
\cite{mcmahon2012the}. As a consequence, the deep investigations of the H-He phase behaviour under relevant planetary conditions are highly desirable \cite{mcmahon2012the,dai2010quantum,wang2013thermophysical,kang2018dynamic,stevenson1975thermodynamics,klepeis1991hydrogen,vorberger2007hydroge,militzer2013,pfaffenzeller1995miscibility,lorenzen2009demixing,li2007benchmarking,chang2023direct}. However, it remains challenging experimentally because the extremes of temperature and pressure are generally difficult to be currently accessible. Previously, the FT-OFDFT has been applied to WDM simulations ( such as H \cite{luo2020towards,kang2020two}, D \cite{luo2020towards,sjostrom2014SD14}, Fe \cite{lambert2006very}, SiO$_2$ \cite{sjostrom2015orbital}, T-D mixture \cite{kang2020unified}, and CHON resin material \cite{zhang2022first}) and becomes routine in recent years. 
The high-quality EOS (less than $\sim$1\%) is required to provide the accurate model to answer some fundamental questions regarding the composition and formation of planets \cite{mcmahon2012the}.  

Here the H, He, and H-He mixtures with two typical densities in the solar interior, 10 $g/cm^3$ and 160 $g/cm^3$ were selected for investigation the matter states of solar radiative zone and core. As shown in Fig.~\ref{fig:MD3}(a) and (b), the EOS curves calculated by FT-OFDFT using XWMF almost overlap with those calculated by FT-KSDFT. It should be stressed that the mean percentage error of pressure ($\overline{{\Delta P^{OF}}/{P^{KS}}} = \frac1N\sum{\frac{|P^{OF}-P^{KS}|}{P^{KS}}}$) was $\simeq 1.4\%$. Due to the superior performance of XWMF in accuracy and generality, FT-OFDFT within XWMF holds great promise for offering a reliable prediction of properties of WDM formed by H and He.

\subsection{Computational cost}
\label{IVC}
\begin{figure}[htp]
\includegraphics[width=0.8 \linewidth]{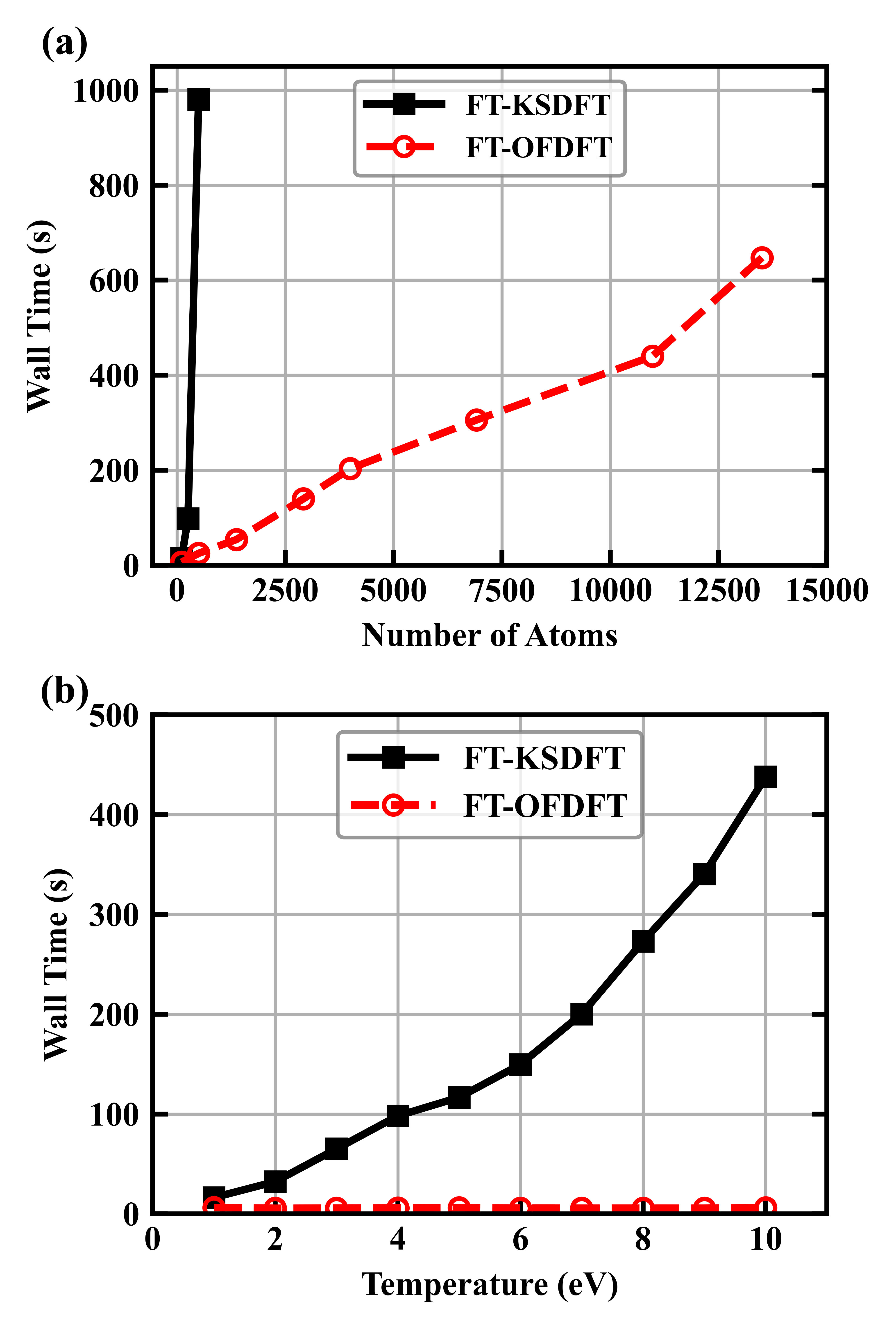}
\caption{\label{timer} Wall times of FT-OFDFT using XWMF and FT-KSDFT for FCC Al supercells static lattice calculations. (a) Wall times as a function of the number of atoms varying from 108 to 13500 at a temperature of 1 eV. (b) Wall times as a function of temperature for 108 atoms supercells.}
\end{figure}
To evaluate the computational efficiency of XWMF, we considered FCC Al supercells with sizes ranging from 108 to 13500 atoms and temperatures ranging from 1 to 10 eV as benchmarks. We compared the wall-time for the static lattice energy calculations using the FT-OFDFT code against that using the FT-KSDFT of VASP 6.1.0 package \cite{Kresse1996vasp1,Kresse1996vasp2}. Note that the projector augmented-wave (PAW) potentials \cite{PAW1994} and gamma-point sampling were employed in FT-KSDFT calculations. All calculations were performed on a node with 2 Intel(R) Xeon(R) Gold 9242 CPUs (48 cores, 2.30 GHz) and 384 GB of RAM. At elevated temperatures, the enough states should be required to accurately describe a thermal ensemble of electrons. Hence, the large computational cost of FT-KSDFT was a major obstacle to practical simulations of WDM. Just as shown in Fig.~\ref{timer}(a) and (b), the wall times of FT-KSDFT grew rapidly as both the number of atoms and temperatures increase. While the wall time of FT-OFDFT using XWMF was quasi-linear scaling with system size [Fig.~\ref{timer}(a)] and it did not scale with temperatures [Fig.~\ref{timer}(b)]. Thus, XWMF method can be used to FT-OFDFT simulations of the WDM with higher temperatures than that accessible in the KSDFT. The dramatic reduction of the wall times with XWMF, as well as its high accuracy, leads us to expect that XWMF is well-suited for simulations of WDM across the wide temperature regime.


\section{Conclusions}
\label{sec:conclusions}
In summary, a nonlocal finite-temperature free-energy energy density functional of XWMF is derived via line integrals and has been systemically applied to several warm dense matters. The accuracy of XWMF is validated by correctly reproducing equations of state and pair distribution functions predicted by FT-KSDFT for various warm dense matters. Remarkably, our findings reveal that the performance of XWMF in accuracy and efficiency is comparable or superior to the existing free-energy density functionals (e.g., WTF, WTF$\beta$vW, and LKTF). Despite its enormous success for expanding the applicability of FT-OFDFT over a very wide temperature range, XWMF derived by uniform electron gas yields the considerable margin of error at lower temperature regime accessible with a conventional KSDFT (less than 1 eV). It is expected that the further improvement of free-energy energy density functional by incorporation of more appropriate electron density response behaviors will be built into the standard tools for investigation of properties of warm dense matters.

\section*{Acknowledgements}
This research was supported by the National Key R\&D Program of China (Grant Nos. 2022YFA1402304, 2023YFA1406200, 2023YFB3003000), the National Natural Science Foundation of China under Grants Nos. T2225013, 12047530, 12274174, 12274171, 12174142, 12034009, 12305002, the Program for JLU Science and Technology Innovative Research Team, and the Science Challenge Project No. TZ2016001. Part of the calculation was performed in the high-performance computing center of Jilin University.

\bibliography{paper}

\begin{thebibliography}{87}%
\makeatletter
\providecommand \@ifxundefined [1]{%
 \@ifx{#1\undefined}
}%
\providecommand \@ifnum [1]{%
 \ifnum #1\expandafter \@firstoftwo
 \else \expandafter \@secondoftwo
 \fi
}%
\providecommand \@ifx [1]{%
 \ifx #1\expandafter \@firstoftwo
 \else \expandafter \@secondoftwo
 \fi
}%
\providecommand \natexlab [1]{#1}%
\providecommand \enquote  [1]{``#1''}%
\providecommand \bibnamefont  [1]{#1}%
\providecommand \bibfnamefont [1]{#1}%
\providecommand \citenamefont [1]{#1}%
\providecommand \href@noop [0]{\@secondoftwo}%
\providecommand \href [0]{\begingroup \@sanitize@url \@href}%
\providecommand \@href[1]{\@@startlink{#1}\@@href}%
\providecommand \@@href[1]{\endgroup#1\@@endlink}%
\providecommand \@sanitize@url [0]{\catcode `\\12\catcode `\$12\catcode
  `\&12\catcode `\#12\catcode `\^12\catcode `\_12\catcode `\%12\relax}%
\providecommand \@@startlink[1]{}%
\providecommand \@@endlink[0]{}%
\providecommand \url  [0]{\begingroup\@sanitize@url \@url }%
\providecommand \@url [1]{\endgroup\@href {#1}{\urlprefix }}%
\providecommand \urlprefix  [0]{URL }%
\providecommand \Eprint [0]{\href }%
\providecommand \doibase [0]{https://doi.org/}%
\providecommand \selectlanguage [0]{\@gobble}%
\providecommand \bibinfo  [0]{\@secondoftwo}%
\providecommand \bibfield  [0]{\@secondoftwo}%
\providecommand \translation [1]{[#1]}%
\providecommand \BibitemOpen [0]{}%
\providecommand \bibitemStop [0]{}%
\providecommand \bibitemNoStop [0]{.\EOS\space}%
\providecommand \EOS [0]{\spacefactor3000\relax}%
\providecommand \BibitemShut  [1]{\csname bibitem#1\endcsname}%
\let\auto@bib@innerbib\@empty
\bibitem [{\citenamefont {Graziani}\ \emph {et~al.}(2014)\citenamefont
  {Graziani}, \citenamefont {Desjarlais}, \citenamefont {Redmer},\ and\
  \citenamefont {Trickey}}]{graziani2014frontiers}%
  \BibitemOpen
  \bibfield  {author} {\bibinfo {author} {\bibfnamefont {F.}~\bibnamefont
  {Graziani}}, \bibinfo {author} {\bibfnamefont {M.~P.}\ \bibnamefont
  {Desjarlais}}, \bibinfo {author} {\bibfnamefont {R.}~\bibnamefont {Redmer}},\
  and\ \bibinfo {author} {\bibfnamefont {S.~B.}\ \bibnamefont {Trickey}},\
  }\href@noop {} {\emph {\bibinfo {title} {Frontiers and challenges in warm
  dense matter}}},\ Vol.~\bibinfo {volume} {96}\ (\bibinfo  {publisher}
  {Springer},\ \bibinfo {year} {2014})\BibitemShut {NoStop}%
\bibitem [{\citenamefont {Riley}(2021)}]{david2021warm}%
  \BibitemOpen
  \bibfield  {author} {\bibinfo {author} {\bibfnamefont {D.}~\bibnamefont
  {Riley}},\ }\href {https://doi.org/10.1088/978-0-7503-2348-2} {\emph
  {\bibinfo {title} {Warm Dense Matter}}}\ (\bibinfo  {publisher} {IOP
  Publishing},\ \bibinfo {year} {2021})\BibitemShut {NoStop}%
\bibitem [{\citenamefont {Hohenberg}\ and\ \citenamefont
  {Kohn}(1964)}]{hk1964}%
  \BibitemOpen
  \bibfield  {author} {\bibinfo {author} {\bibfnamefont {P.}~\bibnamefont
  {Hohenberg}}\ and\ \bibinfo {author} {\bibfnamefont {W.}~\bibnamefont
  {Kohn}},\ }\bibfield  {title} {\bibinfo {title} {Inhomogeneous electron
  gas},\ }\href {https://doi.org/10.1103/PhysRev.136.B864} {\bibfield
  {journal} {\bibinfo  {journal} {Phys. Rev.}\ }\textbf {\bibinfo {volume}
  {136}},\ \bibinfo {pages} {B864} (\bibinfo {year} {1964})}\BibitemShut
  {NoStop}%
\bibitem [{\citenamefont {Kohn}\ and\ \citenamefont
  {Sham}(1965)}]{kohn1965self}%
  \BibitemOpen
  \bibfield  {author} {\bibinfo {author} {\bibfnamefont {W.}~\bibnamefont
  {Kohn}}\ and\ \bibinfo {author} {\bibfnamefont {L.~J.}\ \bibnamefont
  {Sham}},\ }\bibfield  {title} {\bibinfo {title} {Self-consistent equations
  including exchange and correlation effects},\ }\href
  {https://doi.org/10.1103/PhysRev.140.A1133} {\bibfield  {journal} {\bibinfo
  {journal} {Phys. Rev.}\ }\textbf {\bibinfo {volume} {140}},\ \bibinfo {pages}
  {A1133} (\bibinfo {year} {1965})}\BibitemShut {NoStop}%
\bibitem [{\citenamefont {Mermin}(1965)}]{mermin1965thermal}%
  \BibitemOpen
  \bibfield  {author} {\bibinfo {author} {\bibfnamefont {N.~D.}\ \bibnamefont
  {Mermin}},\ }\bibfield  {title} {\bibinfo {title} {Thermal properties of the
  inhomogeneous electron gas},\ }\href
  {https://doi.org/10.1103/PhysRev.137.A1441} {\bibfield  {journal} {\bibinfo
  {journal} {Phys. Rev.}\ }\textbf {\bibinfo {volume} {137}},\ \bibinfo {pages}
  {A1441} (\bibinfo {year} {1965})}\BibitemShut {NoStop}%
\bibitem [{\citenamefont {Parr}\ and\ \citenamefont
  {Yang}(1989)}]{parr1989density}%
  \BibitemOpen
  \bibfield  {author} {\bibinfo {author} {\bibfnamefont {R.}~\bibnamefont
  {Parr}}\ and\ \bibinfo {author} {\bibfnamefont {W.}~\bibnamefont {Yang}},\
  }in\ \href@noop {} {\emph {\bibinfo {booktitle} {Density Functional Theory of
  Atoms and Molecules,}}}\ (\bibinfo  {publisher} {Oxford University Press, New
  York},\ \bibinfo {year} {1989})\BibitemShut {NoStop}%
\bibitem [{\citenamefont {Mazevet}\ \emph {et~al.}(2007)\citenamefont
  {Mazevet}, \citenamefont {Lambert}, \citenamefont {Bottin}, \citenamefont
  {Z\'erah},\ and\ \citenamefont {Cl\'erouin}}]{mazevet2007ab}%
  \BibitemOpen
  \bibfield  {author} {\bibinfo {author} {\bibfnamefont {S.}~\bibnamefont
  {Mazevet}}, \bibinfo {author} {\bibfnamefont {F.}~\bibnamefont {Lambert}},
  \bibinfo {author} {\bibfnamefont {F.}~\bibnamefont {Bottin}}, \bibinfo
  {author} {\bibfnamefont {G.}~\bibnamefont {Z\'erah}},\ and\ \bibinfo {author}
  {\bibfnamefont {J.}~\bibnamefont {Cl\'erouin}},\ }\bibfield  {title}
  {\bibinfo {title} {Ab initio molecular dynamics simulations of dense boron
  plasmas up to the semiclassical thomas-fermi regime},\ }\href
  {https://doi.org/10.1103/PhysRevE.75.056404} {\bibfield  {journal} {\bibinfo
  {journal} {Phys. Rev. E}\ }\textbf {\bibinfo {volume} {75}},\ \bibinfo
  {pages} {056404} (\bibinfo {year} {2007})}\BibitemShut {NoStop}%
\bibitem [{\citenamefont {Holst}\ \emph {et~al.}(2008)\citenamefont {Holst},
  \citenamefont {Redmer},\ and\ \citenamefont
  {Desjarlais}}]{holst2008thermophysical}%
  \BibitemOpen
  \bibfield  {author} {\bibinfo {author} {\bibfnamefont {B.}~\bibnamefont
  {Holst}}, \bibinfo {author} {\bibfnamefont {R.}~\bibnamefont {Redmer}},\ and\
  \bibinfo {author} {\bibfnamefont {M.~P.}\ \bibnamefont {Desjarlais}},\
  }\bibfield  {title} {\bibinfo {title} {Thermophysical properties of warm
  dense hydrogen using quantum molecular dynamics simulations},\ }\href
  {https://doi.org/10.1103/PhysRevB.77.184201} {\bibfield  {journal} {\bibinfo
  {journal} {Phys. Rev. B}\ }\textbf {\bibinfo {volume} {77}},\ \bibinfo
  {pages} {184201} (\bibinfo {year} {2008})}\BibitemShut {NoStop}%
\bibitem [{\citenamefont {Wang}\ and\ \citenamefont
  {Zhang}(2013)}]{wang2013wide}%
  \BibitemOpen
  \bibfield  {author} {\bibinfo {author} {\bibfnamefont {C.}~\bibnamefont
  {Wang}}\ and\ \bibinfo {author} {\bibfnamefont {P.}~\bibnamefont {Zhang}},\
  }\bibfield  {title} {\bibinfo {title} {{Wide range equation of state for
  fluid hydrogen from density functional theory}},\ }\href@noop {} {\bibfield
  {journal} {\bibinfo  {journal} {Phys. Plasmas}\ }\textbf {\bibinfo {volume}
  {20}},\ \bibinfo {pages} {092703} (\bibinfo {year} {2013})}\BibitemShut
  {NoStop}%
\bibitem [{\citenamefont {Kang}\ and\ \citenamefont
  {Dai}(2018)}]{kang2018dynamic}%
  \BibitemOpen
  \bibfield  {author} {\bibinfo {author} {\bibfnamefont {D.}~\bibnamefont
  {Kang}}\ and\ \bibinfo {author} {\bibfnamefont {J.}~\bibnamefont {Dai}},\
  }\bibfield  {title} {\bibinfo {title} {Dynamic electron–ion collisions and
  nuclear quantum effects in quantum simulation of warm dense matter},\ }\href
  {https://doi.org/10.1088/1361-648X/aa9e29} {\bibfield  {journal} {\bibinfo
  {journal} {J. Phys.: Condens. Matter}\ }\textbf {\bibinfo {volume} {30}},\
  \bibinfo {pages} {073002} (\bibinfo {year} {2018})}\BibitemShut {NoStop}%
\bibitem [{\citenamefont {Pollock}\ and\ \citenamefont
  {Ceperley}(1984)}]{pollock1984}%
  \BibitemOpen
  \bibfield  {author} {\bibinfo {author} {\bibfnamefont {E.~L.}\ \bibnamefont
  {Pollock}}\ and\ \bibinfo {author} {\bibfnamefont {D.~M.}\ \bibnamefont
  {Ceperley}},\ }\bibfield  {title} {\bibinfo {title} {Simulation of quantum
  many-body systems by path-integral methods},\ }\href
  {https://doi.org/10.1103/PhysRevB.30.2555} {\bibfield  {journal} {\bibinfo
  {journal} {Phys. Rev. B}\ }\textbf {\bibinfo {volume} {30}},\ \bibinfo
  {pages} {2555} (\bibinfo {year} {1984})}\BibitemShut {NoStop}%
\bibitem [{\citenamefont {Ceperley}(1995)}]{ceperley1995path}%
  \BibitemOpen
  \bibfield  {author} {\bibinfo {author} {\bibfnamefont {D.~M.}\ \bibnamefont
  {Ceperley}},\ }\bibfield  {title} {\bibinfo {title} {Path integrals in the
  theory of condensed helium},\ }\href
  {https://doi.org/10.1103/RevModPhys.67.279} {\bibfield  {journal} {\bibinfo
  {journal} {Rev. Mod. Phys.}\ }\textbf {\bibinfo {volume} {67}},\ \bibinfo
  {pages} {279} (\bibinfo {year} {1995})}\BibitemShut {NoStop}%
\bibitem [{\citenamefont {Driver}\ and\ \citenamefont
  {Militzer}(2012)}]{driver2012all}%
  \BibitemOpen
  \bibfield  {author} {\bibinfo {author} {\bibfnamefont {K.~P.}\ \bibnamefont
  {Driver}}\ and\ \bibinfo {author} {\bibfnamefont {B.}~\bibnamefont
  {Militzer}},\ }\bibfield  {title} {\bibinfo {title} {All-electron path
  integral monte carlo simulations of warm dense matter: Application to water
  and carbon plasmas},\ }\href {https://doi.org/10.1103/PhysRevLett.108.115502}
  {\bibfield  {journal} {\bibinfo  {journal} {Phys. Rev. Lett.}\ }\textbf
  {\bibinfo {volume} {108}},\ \bibinfo {pages} {115502} (\bibinfo {year}
  {2012})}\BibitemShut {NoStop}%
\bibitem [{\citenamefont {Hu}\ \emph {et~al.}(2011)\citenamefont {Hu},
  \citenamefont {Militzer}, \citenamefont {Goncharov},\ and\ \citenamefont
  {Skupsky}}]{hu2011first}%
  \BibitemOpen
  \bibfield  {author} {\bibinfo {author} {\bibfnamefont {S.~X.}\ \bibnamefont
  {Hu}}, \bibinfo {author} {\bibfnamefont {B.}~\bibnamefont {Militzer}},
  \bibinfo {author} {\bibfnamefont {V.~N.}\ \bibnamefont {Goncharov}},\ and\
  \bibinfo {author} {\bibfnamefont {S.}~\bibnamefont {Skupsky}},\ }\bibfield
  {title} {\bibinfo {title} {First-principles equation-of-state table of
  deuterium for inertial confinement fusion applications},\ }\href
  {https://doi.org/10.1103/PhysRevB.84.224109} {\bibfield  {journal} {\bibinfo
  {journal} {Phys. Rev. B}\ }\textbf {\bibinfo {volume} {84}},\ \bibinfo
  {pages} {224109} (\bibinfo {year} {2011})}\BibitemShut {NoStop}%
\bibitem [{\citenamefont {Zhang}\ \emph {et~al.}(2016)\citenamefont {Zhang},
  \citenamefont {Wang}, \citenamefont {Kang}, \citenamefont {Zhang},\ and\
  \citenamefont {He}}]{zhang2016extended}%
  \BibitemOpen
  \bibfield  {author} {\bibinfo {author} {\bibfnamefont {S.}~\bibnamefont
  {Zhang}}, \bibinfo {author} {\bibfnamefont {H.}~\bibnamefont {Wang}},
  \bibinfo {author} {\bibfnamefont {W.}~\bibnamefont {Kang}}, \bibinfo {author}
  {\bibfnamefont {P.}~\bibnamefont {Zhang}},\ and\ \bibinfo {author}
  {\bibfnamefont {X.~T.}\ \bibnamefont {He}},\ }\bibfield  {title} {\bibinfo
  {title} {{Extended application of Kohn-Sham first-principles molecular
  dynamics method with plane wave approximation at high energy—From cold
  materials to hot dense plasmas}},\ }\href {https://doi.org/10.1063/1.4947212}
  {\bibfield  {journal} {\bibinfo  {journal} {Phys. Plasmas}\ }\textbf
  {\bibinfo {volume} {23}},\ \bibinfo {pages} {042707} (\bibinfo {year}
  {2016})}\BibitemShut {NoStop}%
\bibitem [{\citenamefont {Blanchet}\ \emph {et~al.}(2020)\citenamefont
  {Blanchet}, \citenamefont {Torrent},\ and\ \citenamefont
  {Clérouin}}]{blanchet2020requirements}%
  \BibitemOpen
  \bibfield  {author} {\bibinfo {author} {\bibfnamefont {A.}~\bibnamefont
  {Blanchet}}, \bibinfo {author} {\bibfnamefont {M.}~\bibnamefont {Torrent}},\
  and\ \bibinfo {author} {\bibfnamefont {J.}~\bibnamefont {Clérouin}},\
  }\bibfield  {title} {\bibinfo {title} {{Requirements for very high
  temperature Kohn–Sham DFT simulations and how to bypass them}},\ }\href
  {https://doi.org/10.1063/5.0016538} {\bibfield  {journal} {\bibinfo
  {journal} {Phys. Plasmas}\ }\textbf {\bibinfo {volume} {27}},\ \bibinfo
  {pages} {122706} (\bibinfo {year} {2020})}\BibitemShut {NoStop}%
\bibitem [{\citenamefont {Baer}\ \emph {et~al.}(2013)\citenamefont {Baer},
  \citenamefont {Neuhauser},\ and\ \citenamefont {Rabani}}]{baer2013self}%
  \BibitemOpen
  \bibfield  {author} {\bibinfo {author} {\bibfnamefont {R.}~\bibnamefont
  {Baer}}, \bibinfo {author} {\bibfnamefont {D.}~\bibnamefont {Neuhauser}},\
  and\ \bibinfo {author} {\bibfnamefont {E.}~\bibnamefont {Rabani}},\
  }\bibfield  {title} {\bibinfo {title} {Self-averaging stochastic kohn-sham
  density-functional theory},\ }\href
  {https://doi.org/10.1103/PhysRevLett.111.106402} {\bibfield  {journal}
  {\bibinfo  {journal} {Phys. Rev. Lett.}\ }\textbf {\bibinfo {volume} {111}},\
  \bibinfo {pages} {106402} (\bibinfo {year} {2013})}\BibitemShut {NoStop}%
\bibitem [{\citenamefont {White}\ and\ \citenamefont
  {Collins}(2020)}]{white2020fast}%
  \BibitemOpen
  \bibfield  {author} {\bibinfo {author} {\bibfnamefont {A.~J.}\ \bibnamefont
  {White}}\ and\ \bibinfo {author} {\bibfnamefont {L.~A.}\ \bibnamefont
  {Collins}},\ }\bibfield  {title} {\bibinfo {title} {Fast and universal
  kohn-sham density functional theory algorithm for warm dense matter to hot
  dense plasma},\ }\href {https://doi.org/10.1103/PhysRevLett.125.055002}
  {\bibfield  {journal} {\bibinfo  {journal} {Phys. Rev. Lett.}\ }\textbf
  {\bibinfo {volume} {125}},\ \bibinfo {pages} {055002} (\bibinfo {year}
  {2020})}\BibitemShut {NoStop}%
\bibitem [{\citenamefont {Cytter}\ \emph {et~al.}(2018)\citenamefont {Cytter},
  \citenamefont {Rabani}, \citenamefont {Neuhauser},\ and\ \citenamefont
  {Baer}}]{cytter2018stochastic}%
  \BibitemOpen
  \bibfield  {author} {\bibinfo {author} {\bibfnamefont {Y.}~\bibnamefont
  {Cytter}}, \bibinfo {author} {\bibfnamefont {E.}~\bibnamefont {Rabani}},
  \bibinfo {author} {\bibfnamefont {D.}~\bibnamefont {Neuhauser}},\ and\
  \bibinfo {author} {\bibfnamefont {R.}~\bibnamefont {Baer}},\ }\bibfield
  {title} {\bibinfo {title} {Stochastic density functional theory at finite
  temperatures},\ }\href {https://doi.org/10.1103/PhysRevB.97.115207}
  {\bibfield  {journal} {\bibinfo  {journal} {Phys. Rev. B}\ }\textbf {\bibinfo
  {volume} {97}},\ \bibinfo {pages} {115207} (\bibinfo {year}
  {2018})}\BibitemShut {NoStop}%
\bibitem [{\citenamefont {Liu}\ and\ \citenamefont
  {Chen}(2022)}]{liu2022plane}%
  \BibitemOpen
  \bibfield  {author} {\bibinfo {author} {\bibfnamefont {Q.}~\bibnamefont
  {Liu}}\ and\ \bibinfo {author} {\bibfnamefont {M.}~\bibnamefont {Chen}},\
  }\bibfield  {title} {\bibinfo {title} {Plane-wave-based
  stochastic-deterministic density functional theory for extended systems},\
  }\href {https://doi.org/10.1103/PhysRevB.106.125132} {\bibfield  {journal}
  {\bibinfo  {journal} {Phys. Rev. B}\ }\textbf {\bibinfo {volume} {106}},\
  \bibinfo {pages} {125132} (\bibinfo {year} {2022})}\BibitemShut {NoStop}%
\bibitem [{\citenamefont {Wang}\ and\ \citenamefont
  {Carter}(2000)}]{wang2000orbital}%
  \BibitemOpen
  \bibfield  {author} {\bibinfo {author} {\bibfnamefont {Y.~A.}\ \bibnamefont
  {Wang}}\ and\ \bibinfo {author} {\bibfnamefont {E.~A.}\ \bibnamefont
  {Carter}},\ }\href@noop {} {\emph {\bibinfo {title} {Orbital-free
  kinetic-energy density functional theory}}},\ Vol.~\bibinfo {volume} {5}\
  (\bibinfo  {publisher} {Springer},\ \bibinfo {year} {2000})\ pp.\ \bibinfo
  {pages} {117--84}\BibitemShut {NoStop}%
\bibitem [{\citenamefont {Karasiev}\ and\ \citenamefont
  {Trickey}(2012)}]{karasiev2012issues}%
  \BibitemOpen
  \bibfield  {author} {\bibinfo {author} {\bibfnamefont {V.~V.}\ \bibnamefont
  {Karasiev}}\ and\ \bibinfo {author} {\bibfnamefont {S.~B.}\ \bibnamefont
  {Trickey}},\ }\bibfield  {title} {\bibinfo {title} {Issues and challenges in
  orbital-free density functional calculations},\ }\href
  {https://doi.org/https://doi.org/10.1016/j.cpc.2012.06.016} {\bibfield
  {journal} {\bibinfo  {journal} {Comput. Phys. Commun.}\ }\textbf {\bibinfo
  {volume} {183}},\ \bibinfo {pages} {2519} (\bibinfo {year}
  {2012})}\BibitemShut {NoStop}%
\bibitem [{\citenamefont {Karasiev}\ \emph {et~al.}(2012)\citenamefont
  {Karasiev}, \citenamefont {Sjostrom},\ and\ \citenamefont
  {Trickey}}]{Karasiev2012GGA}%
  \BibitemOpen
  \bibfield  {author} {\bibinfo {author} {\bibfnamefont {V.~V.}\ \bibnamefont
  {Karasiev}}, \bibinfo {author} {\bibfnamefont {T.}~\bibnamefont {Sjostrom}},\
  and\ \bibinfo {author} {\bibfnamefont {S.~B.}\ \bibnamefont {Trickey}},\
  }\bibfield  {title} {\bibinfo {title} {Generalized-gradient-approximation
  noninteracting free-energy functionals for orbital-free density functional
  calculations},\ }\href {https://doi.org/10.1103/PhysRevB.86.115101}
  {\bibfield  {journal} {\bibinfo  {journal} {Phys. Rev. B}\ }\textbf {\bibinfo
  {volume} {86}},\ \bibinfo {pages} {115101} (\bibinfo {year}
  {2012})}\BibitemShut {NoStop}%
\bibitem [{\citenamefont {Witt}\ \emph {et~al.}(2018)\citenamefont {Witt},
  \citenamefont {Del~Rio}, \citenamefont {Dieterich},\ and\ \citenamefont
  {Carter}}]{witt2018orbital}%
  \BibitemOpen
  \bibfield  {author} {\bibinfo {author} {\bibfnamefont {W.~C.}\ \bibnamefont
  {Witt}}, \bibinfo {author} {\bibfnamefont {B.~G.}\ \bibnamefont {Del~Rio}},
  \bibinfo {author} {\bibfnamefont {J.~M.}\ \bibnamefont {Dieterich}},\ and\
  \bibinfo {author} {\bibfnamefont {E.~A.}\ \bibnamefont {Carter}},\ }\bibfield
   {title} {\bibinfo {title} {Orbital-free density functional theory for
  materials research},\ }\href {https://doi.org/10.1557/jmr.2017.462}
  {\bibfield  {journal} {\bibinfo  {journal} {J. Mater. Res.}\ }\textbf
  {\bibinfo {volume} {33}},\ \bibinfo {pages} {777} (\bibinfo {year}
  {2018})}\BibitemShut {NoStop}%
\bibitem [{\citenamefont {Mi}\ \emph {et~al.}(2023)\citenamefont {Mi},
  \citenamefont {Luo}, \citenamefont {Trickey},\ and\ \citenamefont
  {Pavanello}}]{mi2023}%
  \BibitemOpen
  \bibfield  {author} {\bibinfo {author} {\bibfnamefont {W.}~\bibnamefont
  {Mi}}, \bibinfo {author} {\bibfnamefont {K.}~\bibnamefont {Luo}}, \bibinfo
  {author} {\bibfnamefont {S.~B.}\ \bibnamefont {Trickey}},\ and\ \bibinfo
  {author} {\bibfnamefont {M.}~\bibnamefont {Pavanello}},\ }\bibfield  {title}
  {\bibinfo {title} {Orbital-free density functional theory: An attractive
  electronic structure method for large-scale first-principles simulations},\
  }\href {https://doi.org/10.1021/acs.chemrev.2c00758} {\bibfield  {journal}
  {\bibinfo  {journal} {Chem. Rev.}\ }\textbf {\bibinfo {volume} {123}},\
  \bibinfo {pages} {12039} (\bibinfo {year} {2023})}\BibitemShut {NoStop}%
\bibitem [{\citenamefont {Fermi}(1927)}]{Fermi1927}%
  \BibitemOpen
  \bibfield  {author} {\bibinfo {author} {\bibfnamefont {E.}~\bibnamefont
  {Fermi}},\ }\bibfield  {title} {\bibinfo {title} {Un metodo statistico per la
  determinazione di alcune priorieta dell'atome},\ }\href@noop {} {\bibfield
  {journal} {\bibinfo  {journal} {Rend. Accad. Naz. Lincei}\ }\textbf {\bibinfo
  {volume} {6}},\ \bibinfo {pages} {32} (\bibinfo {year} {1927})}\BibitemShut
  {NoStop}%
\bibitem [{\citenamefont {Thomas}(1927)}]{Thomas1927}%
  \BibitemOpen
  \bibfield  {author} {\bibinfo {author} {\bibfnamefont {L.~H.}\ \bibnamefont
  {Thomas}},\ }\bibfield  {title} {\bibinfo {title} {The calculation of atomic
  fields},\ }\href@noop {} {\bibfield  {journal} {\bibinfo  {journal} {Math.
  Proc. Cambridge Philos. Soc.}\ }\textbf {\bibinfo {volume} {23}},\ \bibinfo
  {pages} {542} (\bibinfo {year} {1927})}\BibitemShut {NoStop}%
\bibitem [{\citenamefont {Feynman}\ \emph {et~al.}(1949)\citenamefont
  {Feynman}, \citenamefont {Metropolis},\ and\ \citenamefont
  {Teller}}]{feynman1949equations}%
  \BibitemOpen
  \bibfield  {author} {\bibinfo {author} {\bibfnamefont {R.~P.}\ \bibnamefont
  {Feynman}}, \bibinfo {author} {\bibfnamefont {N.}~\bibnamefont
  {Metropolis}},\ and\ \bibinfo {author} {\bibfnamefont {E.}~\bibnamefont
  {Teller}},\ }\bibfield  {title} {\bibinfo {title} {Equations of state of
  elements based on the generalized fermi-thomas theory},\ }\href@noop {}
  {\bibfield  {journal} {\bibinfo  {journal} {Phys. Rev.}\ }\textbf {\bibinfo
  {volume} {75}},\ \bibinfo {pages} {1561} (\bibinfo {year}
  {1949})}\BibitemShut {NoStop}%
\bibitem [{\citenamefont {Bartel}\ \emph {et~al.}(1985)\citenamefont {Bartel},
  \citenamefont {Brack},\ and\ \citenamefont {Durand}}]{bartel1985263extended}%
  \BibitemOpen
  \bibfield  {author} {\bibinfo {author} {\bibfnamefont {J.}~\bibnamefont
  {Bartel}}, \bibinfo {author} {\bibfnamefont {M.}~\bibnamefont {Brack}},\ and\
  \bibinfo {author} {\bibfnamefont {M.}~\bibnamefont {Durand}},\ }\bibfield
  {title} {\bibinfo {title} {Extended thomas-fermi theory at finite
  temperature},\ }\href
  {https://doi.org/https://doi.org/10.1016/0375-9474(85)90071-5} {\bibfield
  {journal} {\bibinfo  {journal} {Nucl. Phys. A}\ }\textbf {\bibinfo {volume}
  {445}},\ \bibinfo {pages} {263} (\bibinfo {year} {1985})}\BibitemShut
  {NoStop}%
\bibitem [{\citenamefont {Perrot}(1979)}]{perrot1979gradient}%
  \BibitemOpen
  \bibfield  {author} {\bibinfo {author} {\bibfnamefont {F.}~\bibnamefont
  {Perrot}},\ }\bibfield  {title} {\bibinfo {title} {Gradient correction to the
  statistical electronic free energy at nonzero temperatures: Application to
  equation-of-state calculations},\ }\href@noop {} {\bibfield  {journal}
  {\bibinfo  {journal} {Phys. Rev. A}\ }\textbf {\bibinfo {volume} {20}},\
  \bibinfo {pages} {586} (\bibinfo {year} {1979})}\BibitemShut {NoStop}%
\bibitem [{\citenamefont {Lambert}\ \emph
  {et~al.}(2006{\natexlab{a}})\citenamefont {Lambert}, \citenamefont
  {Cl\'erouin},\ and\ \citenamefont {Z\'erah}}]{lambert2006very}%
  \BibitemOpen
  \bibfield  {author} {\bibinfo {author} {\bibfnamefont {F.}~\bibnamefont
  {Lambert}}, \bibinfo {author} {\bibfnamefont {J.}~\bibnamefont
  {Cl\'erouin}},\ and\ \bibinfo {author} {\bibfnamefont {G.}~\bibnamefont
  {Z\'erah}},\ }\bibfield  {title} {\bibinfo {title} {Very-high-temperature
  molecular dynamics},\ }\href {https://doi.org/10.1103/PhysRevE.73.016403}
  {\bibfield  {journal} {\bibinfo  {journal} {Phys. Rev. E}\ }\textbf {\bibinfo
  {volume} {73}},\ \bibinfo {pages} {016403} (\bibinfo {year}
  {2006}{\natexlab{a}})}\BibitemShut {NoStop}%
\bibitem [{\citenamefont {Lambert}\ \emph
  {et~al.}(2006{\natexlab{b}})\citenamefont {Lambert}, \citenamefont
  {Clérouin},\ and\ \citenamefont {Mazevet}}]{lambert2006structural}%
  \BibitemOpen
  \bibfield  {author} {\bibinfo {author} {\bibfnamefont {F.}~\bibnamefont
  {Lambert}}, \bibinfo {author} {\bibfnamefont {J.}~\bibnamefont {Clérouin}},\
  and\ \bibinfo {author} {\bibfnamefont {S.}~\bibnamefont {Mazevet}},\
  }\bibfield  {title} {\bibinfo {title} {Structural and dynamical properties of
  hot dense matter by a thomas-fermi-dirac molecular dynamics},\ }\href
  {https://doi.org/10.1209/epl/i2006-10184-7} {\bibfield  {journal} {\bibinfo
  {journal} {Europhys. Lett.}\ }\textbf {\bibinfo {volume} {75}},\ \bibinfo
  {pages} {681} (\bibinfo {year} {2006}{\natexlab{b}})}\BibitemShut {NoStop}%
\bibitem [{\citenamefont {Danel}\ \emph {et~al.}(2006)\citenamefont {Danel},
  \citenamefont {Kazandjian},\ and\ \citenamefont
  {Zérah}}]{danel2006equation}%
  \BibitemOpen
  \bibfield  {author} {\bibinfo {author} {\bibfnamefont {J.-F.}\ \bibnamefont
  {Danel}}, \bibinfo {author} {\bibfnamefont {L.}~\bibnamefont {Kazandjian}},\
  and\ \bibinfo {author} {\bibfnamefont {G.}~\bibnamefont {Zérah}},\
  }\bibfield  {title} {\bibinfo {title} {{Equation of state and sound velocity
  of a helium plasma by Thomas-Fermi-Dirac molecular dynamics}},\ }\href
  {https://doi.org/10.1063/1.2345181} {\bibfield  {journal} {\bibinfo
  {journal} {Phys. Plasmas}\ }\textbf {\bibinfo {volume} {13}},\ \bibinfo
  {pages} {092701} (\bibinfo {year} {2006})}\BibitemShut {NoStop}%
\bibitem [{\citenamefont {Karasiev}\ \emph {et~al.}(2013)\citenamefont
  {Karasiev}, \citenamefont {Chakraborty}, \citenamefont {Shukruto},\ and\
  \citenamefont {Trickey}}]{Karasiev2013VT84F}%
  \BibitemOpen
  \bibfield  {author} {\bibinfo {author} {\bibfnamefont {V.~V.}\ \bibnamefont
  {Karasiev}}, \bibinfo {author} {\bibfnamefont {D.}~\bibnamefont
  {Chakraborty}}, \bibinfo {author} {\bibfnamefont {O.~A.}\ \bibnamefont
  {Shukruto}},\ and\ \bibinfo {author} {\bibfnamefont {S.~B.}\ \bibnamefont
  {Trickey}},\ }\bibfield  {title} {\bibinfo {title} {Nonempirical generalized
  gradient approximation free-energy functional for orbital-free simulations},\
  }\href {https://doi.org/10.1103/PhysRevB.88.161108} {\bibfield  {journal}
  {\bibinfo  {journal} {Phys. Rev. B}\ }\textbf {\bibinfo {volume} {88}},\
  \bibinfo {pages} {161108} (\bibinfo {year} {2013})}\BibitemShut {NoStop}%
\bibitem [{\citenamefont {Luo}\ \emph {et~al.}(2020)\citenamefont {Luo},
  \citenamefont {Karasiev},\ and\ \citenamefont {Trickey}}]{luo2020towards}%
  \BibitemOpen
  \bibfield  {author} {\bibinfo {author} {\bibfnamefont {K.}~\bibnamefont
  {Luo}}, \bibinfo {author} {\bibfnamefont {V.~V.}\ \bibnamefont {Karasiev}},\
  and\ \bibinfo {author} {\bibfnamefont {S.}~\bibnamefont {Trickey}},\
  }\bibfield  {title} {\bibinfo {title} {Towards accurate orbital-free
  simulations: A generalized gradient approximation for the noninteracting free
  energy density functional},\ }\href@noop {} {\bibfield  {journal} {\bibinfo
  {journal} {Phys. Rev. B}\ }\textbf {\bibinfo {volume} {101}},\ \bibinfo
  {pages} {075116} (\bibinfo {year} {2020})}\BibitemShut {NoStop}%
\bibitem [{\citenamefont {Zhang}\ \emph {et~al.}(2022)\citenamefont {Zhang},
  \citenamefont {Karasiev}, \citenamefont {Shaffer}, \citenamefont {Mihaylov},
  \citenamefont {Nichols}, \citenamefont {Paul}, \citenamefont {Goshadze},
  \citenamefont {Ghosh}, \citenamefont {Hinz}, \citenamefont {Epstein},
  \citenamefont {Goedecker},\ and\ \citenamefont {Hu}}]{zhang2022first}%
  \BibitemOpen
  \bibfield  {author} {\bibinfo {author} {\bibfnamefont {S.}~\bibnamefont
  {Zhang}}, \bibinfo {author} {\bibfnamefont {V.~V.}\ \bibnamefont {Karasiev}},
  \bibinfo {author} {\bibfnamefont {N.}~\bibnamefont {Shaffer}}, \bibinfo
  {author} {\bibfnamefont {D.~I.}\ \bibnamefont {Mihaylov}}, \bibinfo {author}
  {\bibfnamefont {K.}~\bibnamefont {Nichols}}, \bibinfo {author} {\bibfnamefont
  {R.}~\bibnamefont {Paul}}, \bibinfo {author} {\bibfnamefont {R.~M.~N.}\
  \bibnamefont {Goshadze}}, \bibinfo {author} {\bibfnamefont {M.}~\bibnamefont
  {Ghosh}}, \bibinfo {author} {\bibfnamefont {J.}~\bibnamefont {Hinz}},
  \bibinfo {author} {\bibfnamefont {R.}~\bibnamefont {Epstein}}, \bibinfo
  {author} {\bibfnamefont {S.}~\bibnamefont {Goedecker}},\ and\ \bibinfo
  {author} {\bibfnamefont {S.~X.}\ \bibnamefont {Hu}},\ }\bibfield  {title}
  {\bibinfo {title} {First-principles equation of state of chon resin for
  inertial confinement fusion applications},\ }\href
  {https://doi.org/10.1103/PhysRevE.106.045207} {\bibfield  {journal} {\bibinfo
   {journal} {Phys. Rev. E}\ }\textbf {\bibinfo {volume} {106}},\ \bibinfo
  {pages} {045207} (\bibinfo {year} {2022})}\BibitemShut {NoStop}%
\bibitem [{\citenamefont {Kang}\ \emph
  {et~al.}(2020{\natexlab{a}})\citenamefont {Kang}, \citenamefont {Luo},
  \citenamefont {Runge},\ and\ \citenamefont {Trickey}}]{kang2020two}%
  \BibitemOpen
  \bibfield  {author} {\bibinfo {author} {\bibfnamefont {D.}~\bibnamefont
  {Kang}}, \bibinfo {author} {\bibfnamefont {K.}~\bibnamefont {Luo}}, \bibinfo
  {author} {\bibfnamefont {K.}~\bibnamefont {Runge}},\ and\ \bibinfo {author}
  {\bibfnamefont {S.~B.}\ \bibnamefont {Trickey}},\ }\bibfield  {title}
  {\bibinfo {title} {{Two-temperature warm dense hydrogen as a test of quantum
  protons driven by orbital-free density functional theory electronic
  forces}},\ }\href {https://doi.org/10.1063/5.0025164} {\bibfield  {journal}
  {\bibinfo  {journal} {Matter Radiat. Extremes}\ }\textbf {\bibinfo {volume}
  {5}},\ \bibinfo {pages} {064403} (\bibinfo {year}
  {2020}{\natexlab{a}})}\BibitemShut {NoStop}%
\bibitem [{\citenamefont {Moldabekov}\ \emph {et~al.}(2021)\citenamefont
  {Moldabekov}, \citenamefont {Dornheim}, \citenamefont {B{\"o}hme},
  \citenamefont {Vorberger},\ and\ \citenamefont
  {Cangi}}]{moldabekov2021relevance}%
  \BibitemOpen
  \bibfield  {author} {\bibinfo {author} {\bibfnamefont {Z.}~\bibnamefont
  {Moldabekov}}, \bibinfo {author} {\bibfnamefont {T.}~\bibnamefont
  {Dornheim}}, \bibinfo {author} {\bibfnamefont {M.}~\bibnamefont {B{\"o}hme}},
  \bibinfo {author} {\bibfnamefont {J.}~\bibnamefont {Vorberger}},\ and\
  \bibinfo {author} {\bibfnamefont {A.}~\bibnamefont {Cangi}},\ }\bibfield
  {title} {\bibinfo {title} {The relevance of electronic perturbations in the
  warm dense electron gas},\ }\href@noop {} {\bibfield  {journal} {\bibinfo
  {journal} {J. Chem. Phys.}\ }\textbf {\bibinfo {volume} {155}},\ \bibinfo
  {pages} {124116} (\bibinfo {year} {2021})}\BibitemShut {NoStop}%
\bibitem [{\citenamefont {Lindhard}(1954)}]{Lindhard1954}%
  \BibitemOpen
  \bibfield  {author} {\bibinfo {author} {\bibfnamefont {J.}~\bibnamefont
  {Lindhard}},\ }\bibfield  {title} {\bibinfo {title} {On the properties of a
  gas of charged particles},\ }\href@noop {} {\bibfield  {journal} {\bibinfo
  {journal} {K. Dan. Vidensk. Selsk. Mat. Fys. Medd.}\ }\textbf {\bibinfo
  {volume} {28}},\ \bibinfo {pages} {1} (\bibinfo {year} {1954})}\BibitemShut
  {NoStop}%
\bibitem [{\citenamefont {Sjostrom}\ and\ \citenamefont
  {Daligault}(2013)}]{sjostrom2013SD13}%
  \BibitemOpen
  \bibfield  {author} {\bibinfo {author} {\bibfnamefont {T.}~\bibnamefont
  {Sjostrom}}\ and\ \bibinfo {author} {\bibfnamefont {J.}~\bibnamefont
  {Daligault}},\ }\bibfield  {title} {\bibinfo {title} {Nonlocal orbital-free
  noninteracting free-energy functional for warm dense matter},\ }\href
  {https://doi.org/10.1103/PhysRevB.88.195103} {\bibfield  {journal} {\bibinfo
  {journal} {Phys. Rev. B}\ }\textbf {\bibinfo {volume} {88}},\ \bibinfo
  {pages} {195103} (\bibinfo {year} {2013})}\BibitemShut {NoStop}%
\bibitem [{\citenamefont {Moldabekov}\ \emph
  {et~al.}(2023{\natexlab{a}})\citenamefont {Moldabekov}, \citenamefont {Shao},
  \citenamefont {Pavanello}, \citenamefont {Vorberger}, \citenamefont
  {Graziani},\ and\ \citenamefont {Dornheim}}]{moldabekov2023imposing}%
  \BibitemOpen
  \bibfield  {author} {\bibinfo {author} {\bibfnamefont {Z.~A.}\ \bibnamefont
  {Moldabekov}}, \bibinfo {author} {\bibfnamefont {X.}~\bibnamefont {Shao}},
  \bibinfo {author} {\bibfnamefont {M.}~\bibnamefont {Pavanello}}, \bibinfo
  {author} {\bibfnamefont {J.}~\bibnamefont {Vorberger}}, \bibinfo {author}
  {\bibfnamefont {F.}~\bibnamefont {Graziani}},\ and\ \bibinfo {author}
  {\bibfnamefont {T.}~\bibnamefont {Dornheim}},\ }\bibfield  {title} {\bibinfo
  {title} {Imposing correct jellium response is key to predict the density
  response by orbital-free dft},\ }\href
  {https://doi.org/10.1103/PhysRevB.108.235168} {\bibfield  {journal} {\bibinfo
   {journal} {Phys. Rev. B}\ }\textbf {\bibinfo {volume} {108}},\ \bibinfo
  {pages} {235168} (\bibinfo {year} {2023}{\natexlab{a}})}\BibitemShut
  {NoStop}%
\bibitem [{\citenamefont {Wang}\ and\ \citenamefont
  {Teter}(1992)}]{wang1992kinetic}%
  \BibitemOpen
  \bibfield  {author} {\bibinfo {author} {\bibfnamefont {L.-W.}\ \bibnamefont
  {Wang}}\ and\ \bibinfo {author} {\bibfnamefont {M.~P.}\ \bibnamefont
  {Teter}},\ }\bibfield  {title} {\bibinfo {title} {Kinetic-energy functional
  of the electron density},\ }\href@noop {} {\bibfield  {journal} {\bibinfo
  {journal} {Phys. Rev. B}\ }\textbf {\bibinfo {volume} {45}},\ \bibinfo
  {pages} {13196} (\bibinfo {year} {1992})}\BibitemShut {NoStop}%
\bibitem [{\citenamefont {Giuliani}\ and\ \citenamefont
  {Vignale}(2008)}]{giuliani2008quantum}%
  \BibitemOpen
  \bibfield  {author} {\bibinfo {author} {\bibfnamefont {G.}~\bibnamefont
  {Giuliani}}\ and\ \bibinfo {author} {\bibfnamefont {G.}~\bibnamefont
  {Vignale}},\ }\href@noop {} {\emph {\bibinfo {title} {Quantum theory of the
  electron liquid}}}\ (\bibinfo  {publisher} {Cambridge university press},\
  \bibinfo {year} {2008})\BibitemShut {NoStop}%
\bibitem [{\citenamefont {Sjostrom}\ and\ \citenamefont
  {Daligault}(2014)}]{sjostrom2014SD14}%
  \BibitemOpen
  \bibfield  {author} {\bibinfo {author} {\bibfnamefont {T.}~\bibnamefont
  {Sjostrom}}\ and\ \bibinfo {author} {\bibfnamefont {J.}~\bibnamefont
  {Daligault}},\ }\bibfield  {title} {\bibinfo {title} {Fast and accurate
  quantum molecular dynamics of dense plasmas across temperature regimes},\
  }\href {https://doi.org/10.1103/PhysRevLett.113.155006} {\bibfield  {journal}
  {\bibinfo  {journal} {Phys. Rev. Lett.}\ }\textbf {\bibinfo {volume} {113}},\
  \bibinfo {pages} {155006} (\bibinfo {year} {2014})}\BibitemShut {NoStop}%
\bibitem [{\citenamefont {Xu}\ \emph {et~al.}(2019)\citenamefont {Xu},
  \citenamefont {Wang},\ and\ \citenamefont {Ma}}]{xu2019nonlocal}%
  \BibitemOpen
  \bibfield  {author} {\bibinfo {author} {\bibfnamefont {Q.}~\bibnamefont
  {Xu}}, \bibinfo {author} {\bibfnamefont {Y.}~\bibnamefont {Wang}},\ and\
  \bibinfo {author} {\bibfnamefont {Y.}~\bibnamefont {Ma}},\ }\bibfield
  {title} {\bibinfo {title} {Nonlocal kinetic energy density functional via
  line integrals and its application to orbital-free density functional
  theory},\ }\href@noop {} {\bibfield  {journal} {\bibinfo  {journal} {Phys.
  Rev. B}\ }\textbf {\bibinfo {volume} {100}},\ \bibinfo {pages} {205132}
  (\bibinfo {year} {2019})}\BibitemShut {NoStop}%
\bibitem [{\citenamefont {Karasiev}\ \emph {et~al.}(2014)\citenamefont
  {Karasiev}, \citenamefont {Sjostrom}, \citenamefont {Dufty},\ and\
  \citenamefont {Trickey}}]{karasiev2014accurate}%
  \BibitemOpen
  \bibfield  {author} {\bibinfo {author} {\bibfnamefont {V.~V.}\ \bibnamefont
  {Karasiev}}, \bibinfo {author} {\bibfnamefont {T.}~\bibnamefont {Sjostrom}},
  \bibinfo {author} {\bibfnamefont {J.}~\bibnamefont {Dufty}},\ and\ \bibinfo
  {author} {\bibfnamefont {S.}~\bibnamefont {Trickey}},\ }\bibfield  {title}
  {\bibinfo {title} {Accurate homogeneous electron gas exchange-correlation
  free energy for local spin-density calculations},\ }\href@noop {} {\bibfield
  {journal} {\bibinfo  {journal} {Phys. Rev. Lett.}\ }\textbf {\bibinfo
  {volume} {112}},\ \bibinfo {pages} {076403} (\bibinfo {year}
  {2014})}\BibitemShut {NoStop}%
\bibitem [{\citenamefont {Karasiev}\ \emph {et~al.}(2018)\citenamefont
  {Karasiev}, \citenamefont {Dufty},\ and\ \citenamefont
  {Trickey}}]{karasiev2018nonempirical}%
  \BibitemOpen
  \bibfield  {author} {\bibinfo {author} {\bibfnamefont {V.~V.}\ \bibnamefont
  {Karasiev}}, \bibinfo {author} {\bibfnamefont {J.~W.}\ \bibnamefont
  {Dufty}},\ and\ \bibinfo {author} {\bibfnamefont {S.}~\bibnamefont
  {Trickey}},\ }\bibfield  {title} {\bibinfo {title} {Nonempirical semilocal
  free-energy density functional for matter under extreme conditions},\
  }\href@noop {} {\bibfield  {journal} {\bibinfo  {journal} {Phys. Rev. Lett.}\
  }\textbf {\bibinfo {volume} {120}},\ \bibinfo {pages} {076401} (\bibinfo
  {year} {2018})}\BibitemShut {NoStop}%
\bibitem [{\citenamefont {Mihaylov}\ \emph {et~al.}(2020)\citenamefont
  {Mihaylov}, \citenamefont {Karasiev},\ and\ \citenamefont
  {Hu}}]{mihaylov2020thermal}%
  \BibitemOpen
  \bibfield  {author} {\bibinfo {author} {\bibfnamefont {D.~I.}\ \bibnamefont
  {Mihaylov}}, \bibinfo {author} {\bibfnamefont {V.~V.}\ \bibnamefont
  {Karasiev}},\ and\ \bibinfo {author} {\bibfnamefont {S.~X.}\ \bibnamefont
  {Hu}},\ }\bibfield  {title} {\bibinfo {title} {Thermal hybrid
  exchange-correlation density functional for improving the description of warm
  dense matter},\ }\href {https://doi.org/10.1103/PhysRevB.101.245141}
  {\bibfield  {journal} {\bibinfo  {journal} {Phys. Rev. B}\ }\textbf {\bibinfo
  {volume} {101}},\ \bibinfo {pages} {245141} (\bibinfo {year}
  {2020})}\BibitemShut {NoStop}%
\bibitem [{\citenamefont {Karasiev}\ \emph {et~al.}(2022)\citenamefont
  {Karasiev}, \citenamefont {Mihaylov},\ and\ \citenamefont
  {Hu}}]{karasiev2022meta}%
  \BibitemOpen
  \bibfield  {author} {\bibinfo {author} {\bibfnamefont {V.~V.}\ \bibnamefont
  {Karasiev}}, \bibinfo {author} {\bibfnamefont {D.}~\bibnamefont {Mihaylov}},\
  and\ \bibinfo {author} {\bibfnamefont {S.}~\bibnamefont {Hu}},\ }\bibfield
  {title} {\bibinfo {title} {Meta-gga exchange-correlation free energy density
  functional to increase the accuracy of warm dense matter simulations},\
  }\href@noop {} {\bibfield  {journal} {\bibinfo  {journal} {Phys. Rev. B}\
  }\textbf {\bibinfo {volume} {105}},\ \bibinfo {pages} {L081109} (\bibinfo
  {year} {2022})}\BibitemShut {NoStop}%
\bibitem [{\citenamefont {Moldabekov}\ \emph {et~al.}(2022)\citenamefont
  {Moldabekov}, \citenamefont {Dornheim}, \citenamefont {Vorberger},\ and\
  \citenamefont {Cangi}}]{moldabekov2022benchmarking}%
  \BibitemOpen
  \bibfield  {author} {\bibinfo {author} {\bibfnamefont {Z.}~\bibnamefont
  {Moldabekov}}, \bibinfo {author} {\bibfnamefont {T.}~\bibnamefont
  {Dornheim}}, \bibinfo {author} {\bibfnamefont {J.}~\bibnamefont
  {Vorberger}},\ and\ \bibinfo {author} {\bibfnamefont {A.}~\bibnamefont
  {Cangi}},\ }\bibfield  {title} {\bibinfo {title} {Benchmarking
  exchange-correlation functionals in the spin-polarized inhomogeneous electron
  gas under warm dense conditions},\ }\href@noop {} {\bibfield  {journal}
  {\bibinfo  {journal} {Phys. Rev. B}\ }\textbf {\bibinfo {volume} {105}},\
  \bibinfo {pages} {035134} (\bibinfo {year} {2022})}\BibitemShut {NoStop}%
\bibitem [{\citenamefont {Moldabekov}\ \emph
  {et~al.}(2023{\natexlab{b}})\citenamefont {Moldabekov}, \citenamefont
  {Lokamani}, \citenamefont {Vorberger}, \citenamefont {Cangi},\ and\
  \citenamefont {Dornheim}}]{moldabekov2023assessing}%
  \BibitemOpen
  \bibfield  {author} {\bibinfo {author} {\bibfnamefont {Z.~A.}\ \bibnamefont
  {Moldabekov}}, \bibinfo {author} {\bibfnamefont {M.}~\bibnamefont
  {Lokamani}}, \bibinfo {author} {\bibfnamefont {J.}~\bibnamefont {Vorberger}},
  \bibinfo {author} {\bibfnamefont {A.}~\bibnamefont {Cangi}},\ and\ \bibinfo
  {author} {\bibfnamefont {T.}~\bibnamefont {Dornheim}},\ }\bibfield  {title}
  {\bibinfo {title} {Assessing the accuracy of hybrid exchange-correlation
  functionals for the density response of warm dense electrons},\ }\href@noop
  {} {\bibfield  {journal} {\bibinfo  {journal} {J. Chem. Phys.}\ }\textbf
  {\bibinfo {volume} {158}},\ \bibinfo {pages} {094105} (\bibinfo {year}
  {2023}{\natexlab{b}})}\BibitemShut {NoStop}%
\bibitem [{\citenamefont {Karasiev}\ \emph {et~al.}(2015)\citenamefont
  {Karasiev}, \citenamefont {Chakraborty},\ and\ \citenamefont
  {Trickey}}]{valentin2015improved}%
  \BibitemOpen
  \bibfield  {author} {\bibinfo {author} {\bibfnamefont {V.~V.}\ \bibnamefont
  {Karasiev}}, \bibinfo {author} {\bibfnamefont {D.}~\bibnamefont
  {Chakraborty}},\ and\ \bibinfo {author} {\bibfnamefont {S.}~\bibnamefont
  {Trickey}},\ }\bibfield  {title} {\bibinfo {title} {Improved analytical
  representation of combinations of fermi–dirac integrals for
  finite-temperature density functional calculations},\ }\href
  {https://doi.org/https://doi.org/10.1016/j.cpc.2015.03.002} {\bibfield
  {journal} {\bibinfo  {journal} {Comput. Phys. Commun.}\ }\textbf {\bibinfo
  {volume} {192}},\ \bibinfo {pages} {114} (\bibinfo {year}
  {2015})}\BibitemShut {NoStop}%
\bibitem [{\citenamefont {Weizs{\"a}cker}(1935)}]{weizsacker1935}%
  \BibitemOpen
  \bibfield  {author} {\bibinfo {author} {\bibfnamefont {C.~F.~v.}\
  \bibnamefont {Weizs{\"a}cker}},\ }\bibfield  {title} {\bibinfo {title} {Zur
  theorie der kernmassen},\ }\href {https://doi.org/10.1007/BF01337700}
  {\bibfield  {journal} {\bibinfo  {journal} {Z. Phys.}\ }\textbf {\bibinfo
  {volume} {96}},\ \bibinfo {pages} {431} (\bibinfo {year} {1935})}\BibitemShut
  {NoStop}%
\bibitem [{\citenamefont {Chai}\ and\ \citenamefont
  {Weeks}(2007)}]{chai2007orbital}%
  \BibitemOpen
  \bibfield  {author} {\bibinfo {author} {\bibfnamefont {J.-D.}\ \bibnamefont
  {Chai}}\ and\ \bibinfo {author} {\bibfnamefont {J.~D.}\ \bibnamefont
  {Weeks}},\ }\bibfield  {title} {\bibinfo {title} {Orbital-free density
  functional theory: Kinetic potentials and ab initio local pseudopotentials},\
  }\href {https://doi.org/10.1103/PhysRevB.75.205122} {\bibfield  {journal}
  {\bibinfo  {journal} {Phys. Rev. B}\ }\textbf {\bibinfo {volume} {75}},\
  \bibinfo {pages} {205122} (\bibinfo {year} {2007})}\BibitemShut {NoStop}%
\bibitem [{Note111()}]{Note111}%
  \BibitemOpen
  \bibinfo {note} {See Supplemental Material at [URL will be inserted by
  publisher]}\BibitemShut {NoStop}%
\bibitem [{\citenamefont {Xu}\ \emph {et~al.}(2022)\citenamefont {Xu},
  \citenamefont {Ma}, \citenamefont {Mi}, \citenamefont {Wang},\ and\
  \citenamefont {Ma}}]{xu2022nonlocal}%
  \BibitemOpen
  \bibfield  {author} {\bibinfo {author} {\bibfnamefont {Q.}~\bibnamefont
  {Xu}}, \bibinfo {author} {\bibfnamefont {C.}~\bibnamefont {Ma}}, \bibinfo
  {author} {\bibfnamefont {W.}~\bibnamefont {Mi}}, \bibinfo {author}
  {\bibfnamefont {Y.}~\bibnamefont {Wang}},\ and\ \bibinfo {author}
  {\bibfnamefont {Y.}~\bibnamefont {Ma}},\ }\bibfield  {title} {\bibinfo
  {title} {Nonlocal pseudopotential energy density functional for orbital-free
  density functional theory},\ }\href@noop {} {\bibfield  {journal} {\bibinfo
  {journal} {Nat. Commun.}\ }\textbf {\bibinfo {volume} {13}},\ \bibinfo
  {pages} {1385} (\bibinfo {year} {2022})}\BibitemShut {NoStop}%
\bibitem [{\citenamefont {Mi}\ \emph {et~al.}(2016)\citenamefont {Mi},
  \citenamefont {Shao}, \citenamefont {Su}, \citenamefont {Zhou}, \citenamefont
  {Zhang}, \citenamefont {Li}, \citenamefont {Wang}, \citenamefont {Zhang},
  \citenamefont {Miao}, \citenamefont {Wang},\ and\ \citenamefont
  {Ma}}]{mi2016atlas}%
  \BibitemOpen
  \bibfield  {author} {\bibinfo {author} {\bibfnamefont {W.}~\bibnamefont
  {Mi}}, \bibinfo {author} {\bibfnamefont {X.}~\bibnamefont {Shao}}, \bibinfo
  {author} {\bibfnamefont {C.}~\bibnamefont {Su}}, \bibinfo {author}
  {\bibfnamefont {Y.}~\bibnamefont {Zhou}}, \bibinfo {author} {\bibfnamefont
  {S.}~\bibnamefont {Zhang}}, \bibinfo {author} {\bibfnamefont
  {Q.}~\bibnamefont {Li}}, \bibinfo {author} {\bibfnamefont {H.}~\bibnamefont
  {Wang}}, \bibinfo {author} {\bibfnamefont {L.}~\bibnamefont {Zhang}},
  \bibinfo {author} {\bibfnamefont {M.}~\bibnamefont {Miao}}, \bibinfo {author}
  {\bibfnamefont {Y.}~\bibnamefont {Wang}},\ and\ \bibinfo {author}
  {\bibfnamefont {Y.}~\bibnamefont {Ma}},\ }\bibfield  {title} {\bibinfo
  {title} {{ATLAS:} a real-space finite-difference implementation of
  orbital-free density functional theory},\ }\href
  {https://doi.org/https://doi.org/10.1016/j.cpc.2015.11.004} {\bibfield
  {journal} {\bibinfo  {journal} {Comput. Phys. Commun.}\ }\textbf {\bibinfo
  {volume} {200}},\ \bibinfo {pages} {87} (\bibinfo {year} {2016})}\BibitemShut
  {NoStop}%
\bibitem [{\citenamefont {Shao}\ \emph {et~al.}(2018)\citenamefont {Shao},
  \citenamefont {Xu}, \citenamefont {Wang}, \citenamefont {Lv}, \citenamefont
  {Wang},\ and\ \citenamefont {Ma}}]{shao2018large}%
  \BibitemOpen
  \bibfield  {author} {\bibinfo {author} {\bibfnamefont {X.}~\bibnamefont
  {Shao}}, \bibinfo {author} {\bibfnamefont {Q.}~\bibnamefont {Xu}}, \bibinfo
  {author} {\bibfnamefont {S.}~\bibnamefont {Wang}}, \bibinfo {author}
  {\bibfnamefont {J.}~\bibnamefont {Lv}}, \bibinfo {author} {\bibfnamefont
  {Y.}~\bibnamefont {Wang}},\ and\ \bibinfo {author} {\bibfnamefont
  {Y.}~\bibnamefont {Ma}},\ }\bibfield  {title} {\bibinfo {title} {Large-scale
  ab initio simulations for periodic system},\ }\href@noop {} {\bibfield
  {journal} {\bibinfo  {journal} {Comput. Phys. Commun.}\ }\textbf {\bibinfo
  {volume} {233}},\ \bibinfo {pages} {78} (\bibinfo {year} {2018})}\BibitemShut
  {NoStop}%
\bibitem [{\citenamefont {Perdew}\ and\ \citenamefont
  {Zunger}(1981)}]{Perdew1981PZ}%
  \BibitemOpen
  \bibfield  {author} {\bibinfo {author} {\bibfnamefont {J.~P.}\ \bibnamefont
  {Perdew}}\ and\ \bibinfo {author} {\bibfnamefont {A.}~\bibnamefont
  {Zunger}},\ }\bibfield  {title} {\bibinfo {title} {Self-interaction
  correction to density-functional approximations for many-electron systems},\
  }\href {https://doi.org/10.1103/PhysRevB.23.5048} {\bibfield  {journal}
  {\bibinfo  {journal} {Phys. Rev. B}\ }\textbf {\bibinfo {volume} {23}},\
  \bibinfo {pages} {5048} (\bibinfo {year} {1981})}\BibitemShut {NoStop}%
\bibitem [{\citenamefont {Zhou}\ \emph {et~al.}(2004)\citenamefont {Zhou},
  \citenamefont {Alexander~Wang},\ and\ \citenamefont
  {Carter}}]{Zhou2004BLPS1}%
  \BibitemOpen
  \bibfield  {author} {\bibinfo {author} {\bibfnamefont {B.}~\bibnamefont
  {Zhou}}, \bibinfo {author} {\bibfnamefont {Y.}~\bibnamefont
  {Alexander~Wang}},\ and\ \bibinfo {author} {\bibfnamefont {E.~A.}\
  \bibnamefont {Carter}},\ }\bibfield  {title} {\bibinfo {title} {Transferable
  local pseudopotentials derived via inversion of the {Kohn-Sham} equations in
  a bulk environment},\ }\href {https://doi.org/10.1103/PhysRevB.69.125109}
  {\bibfield  {journal} {\bibinfo  {journal} {Phys. Rev. B}\ }\textbf {\bibinfo
  {volume} {69}},\ \bibinfo {pages} {125109} (\bibinfo {year}
  {2004})}\BibitemShut {NoStop}%
\bibitem [{\citenamefont {Huang}\ and\ \citenamefont
  {Carter}(2008)}]{Huang2008BLPS2}%
  \BibitemOpen
  \bibfield  {author} {\bibinfo {author} {\bibfnamefont {C.}~\bibnamefont
  {Huang}}\ and\ \bibinfo {author} {\bibfnamefont {E.~A.}\ \bibnamefont
  {Carter}},\ }\bibfield  {title} {\bibinfo {title} {Transferable local
  pseudopotentials for magnesium{,} aluminum and silicon},\ }\href
  {https://doi.org/10.1039/B810407G} {\bibfield  {journal} {\bibinfo  {journal}
  {Phys. Chem. Chem. Phys.}\ }\textbf {\bibinfo {volume} {10}},\ \bibinfo
  {pages} {7109} (\bibinfo {year} {2008})}\BibitemShut {NoStop}%
\bibitem [{\citenamefont {Heine}\ and\ \citenamefont
  {Abarenkov}(1964)}]{heine1964new}%
  \BibitemOpen
  \bibfield  {author} {\bibinfo {author} {\bibfnamefont {V.}~\bibnamefont
  {Heine}}\ and\ \bibinfo {author} {\bibfnamefont {I.~V.}\ \bibnamefont
  {Abarenkov}},\ }\bibfield  {title} {\bibinfo {title} {A new method for the
  electronic structure of metals},\ }\href@noop {} {\bibfield  {journal}
  {\bibinfo  {journal} {Philos. Mag.}\ }\textbf {\bibinfo {volume} {9}},\
  \bibinfo {pages} {451} (\bibinfo {year} {1964})}\BibitemShut {NoStop}%
\bibitem [{\citenamefont {Goodwin}\ \emph {et~al.}(1990)\citenamefont
  {Goodwin}, \citenamefont {Needs},\ and\ \citenamefont
  {Heine}}]{goodwin1990pseudopotential}%
  \BibitemOpen
  \bibfield  {author} {\bibinfo {author} {\bibfnamefont {L.}~\bibnamefont
  {Goodwin}}, \bibinfo {author} {\bibfnamefont {R.}~\bibnamefont {Needs}},\
  and\ \bibinfo {author} {\bibfnamefont {V.}~\bibnamefont {Heine}},\ }\bibfield
   {title} {\bibinfo {title} {A pseudopotential total energy study of
  impurity-promoted intergranular embrittlement},\ }\href@noop {} {\bibfield
  {journal} {\bibinfo  {journal} {J. Phys.: Condens. Matter}\ }\textbf
  {\bibinfo {volume} {2}},\ \bibinfo {pages} {351} (\bibinfo {year}
  {1990})}\BibitemShut {NoStop}%
\bibitem [{\citenamefont {Clark}\ \emph {et~al.}(2005)\citenamefont {Clark},
  \citenamefont {Segall}, \citenamefont {Pickard}, \citenamefont {Hasnip},
  \citenamefont {Probert}, \citenamefont {Refson},\ and\ \citenamefont
  {Payne}}]{CASTEP2005}%
  \BibitemOpen
  \bibfield  {author} {\bibinfo {author} {\bibfnamefont {S.~J.}\ \bibnamefont
  {Clark}}, \bibinfo {author} {\bibfnamefont {M.~D.}\ \bibnamefont {Segall}},
  \bibinfo {author} {\bibfnamefont {C.~J.}\ \bibnamefont {Pickard}}, \bibinfo
  {author} {\bibfnamefont {P.~J.}\ \bibnamefont {Hasnip}}, \bibinfo {author}
  {\bibfnamefont {M.~J.}\ \bibnamefont {Probert}}, \bibinfo {author}
  {\bibfnamefont {K.}~\bibnamefont {Refson}},\ and\ \bibinfo {author}
  {\bibfnamefont {M.}~\bibnamefont {Payne}},\ }\bibfield  {title} {\bibinfo
  {title} {First principles methods using {CASTEP}},\ }\href
  {https://doi.org/doi:10.1524/zkri.220.5.567.65075} {\bibfield  {journal}
  {\bibinfo  {journal} {Z. Kristall.}\ }\textbf {\bibinfo {volume} {220}},\
  \bibinfo {pages} {567} (\bibinfo {year} {2005})}\BibitemShut {NoStop}%
\bibitem [{\citenamefont {Monkhorst}\ and\ \citenamefont
  {Pack}(1976)}]{Monkhorst1976}%
  \BibitemOpen
  \bibfield  {author} {\bibinfo {author} {\bibfnamefont {H.~J.}\ \bibnamefont
  {Monkhorst}}\ and\ \bibinfo {author} {\bibfnamefont {J.~D.}\ \bibnamefont
  {Pack}},\ }\bibfield  {title} {\bibinfo {title} {Special points for
  brillouin-zone integrations},\ }\href
  {https://doi.org/10.1103/PhysRevB.13.5188} {\bibfield  {journal} {\bibinfo
  {journal} {Phys. Rev. B}\ }\textbf {\bibinfo {volume} {13}},\ \bibinfo
  {pages} {5188} (\bibinfo {year} {1976})}\BibitemShut {NoStop}%
\bibitem [{\citenamefont {Troullier}\ and\ \citenamefont
  {Martins}(1991)}]{Troullier1991}%
  \BibitemOpen
  \bibfield  {author} {\bibinfo {author} {\bibfnamefont {N.}~\bibnamefont
  {Troullier}}\ and\ \bibinfo {author} {\bibfnamefont {J.~L.}\ \bibnamefont
  {Martins}},\ }\bibfield  {title} {\bibinfo {title} {Efficient
  pseudopotentials for plane-wave calculations},\ }\href
  {https://doi.org/10.1103/PhysRevB.43.1993} {\bibfield  {journal} {\bibinfo
  {journal} {Phys. Rev. B}\ }\textbf {\bibinfo {volume} {43}},\ \bibinfo
  {pages} {1993} (\bibinfo {year} {1991})}\BibitemShut {NoStop}%
\bibitem [{\citenamefont {Fuchs}\ and\ \citenamefont
  {Scheffler}(1999)}]{fuchs1999fhi98}%
  \BibitemOpen
  \bibfield  {author} {\bibinfo {author} {\bibfnamefont {M.}~\bibnamefont
  {Fuchs}}\ and\ \bibinfo {author} {\bibfnamefont {M.}~\bibnamefont
  {Scheffler}},\ }\bibfield  {title} {\bibinfo {title} {Ab initio
  pseudopotentials for electronic structure calculations of poly-atomic systems
  using density-functional theory},\ }\href
  {https://doi.org/https://doi.org/10.1016/S0010-4655(98)00201-X} {\bibfield
  {journal} {\bibinfo  {journal} {Comput. Phys. Commun.}\ }\textbf {\bibinfo
  {volume} {119}},\ \bibinfo {pages} {67} (\bibinfo {year} {1999})}\BibitemShut
  {NoStop}%
\bibitem [{\citenamefont {Chen}\ \emph {et~al.}(2010)\citenamefont {Chen},
  \citenamefont {Guo},\ and\ \citenamefont {He}}]{chen2010abacus}%
  \BibitemOpen
  \bibfield  {author} {\bibinfo {author} {\bibfnamefont {M.}~\bibnamefont
  {Chen}}, \bibinfo {author} {\bibfnamefont {G.-C.}\ \bibnamefont {Guo}},\ and\
  \bibinfo {author} {\bibfnamefont {L.}~\bibnamefont {He}},\ }\bibfield
  {title} {\bibinfo {title} {Systematically improvable optimized atomic basis
  sets for ab initio calculations},\ }\href
  {https://doi.org/10.1088/0953-8984/22/44/445501} {\bibfield  {journal}
  {\bibinfo  {journal} {J. Phys.: Condens. Matter}\ }\textbf {\bibinfo {volume}
  {22}},\ \bibinfo {pages} {445501} (\bibinfo {year} {2010})}\BibitemShut
  {NoStop}%
\bibitem [{\citenamefont {Li}\ \emph {et~al.}(2016)\citenamefont {Li},
  \citenamefont {Liu}, \citenamefont {Chen}, \citenamefont {Lin}, \citenamefont
  {Ren}, \citenamefont {Lin}, \citenamefont {Yang},\ and\ \citenamefont
  {He}}]{li2016large}%
  \BibitemOpen
  \bibfield  {author} {\bibinfo {author} {\bibfnamefont {P.}~\bibnamefont
  {Li}}, \bibinfo {author} {\bibfnamefont {X.}~\bibnamefont {Liu}}, \bibinfo
  {author} {\bibfnamefont {M.}~\bibnamefont {Chen}}, \bibinfo {author}
  {\bibfnamefont {P.}~\bibnamefont {Lin}}, \bibinfo {author} {\bibfnamefont
  {X.}~\bibnamefont {Ren}}, \bibinfo {author} {\bibfnamefont {L.}~\bibnamefont
  {Lin}}, \bibinfo {author} {\bibfnamefont {C.}~\bibnamefont {Yang}},\ and\
  \bibinfo {author} {\bibfnamefont {L.}~\bibnamefont {He}},\ }\bibfield
  {title} {\bibinfo {title} {Large-scale ab initio simulations based on
  systematically improvable atomic basis},\ }\href
  {https://doi.org/https://doi.org/10.1016/j.commatsci.2015.07.004} {\bibfield
  {journal} {\bibinfo  {journal} {Comput. Mater. Sci.}\ }\textbf {\bibinfo
  {volume} {112}},\ \bibinfo {pages} {503} (\bibinfo {year} {2016})},\ \bibinfo
  {note} {computational Materials Science in China}\BibitemShut {NoStop}%
\bibitem [{\citenamefont {Nosé}(1984)}]{nose1984a}%
  \BibitemOpen
  \bibfield  {author} {\bibinfo {author} {\bibfnamefont {S.}~\bibnamefont
  {Nosé}},\ }\bibfield  {title} {\bibinfo {title} {{A unified formulation of
  the constant temperature molecular dynamics methods}},\ }\href
  {https://doi.org/10.1063/1.447334} {\bibfield  {journal} {\bibinfo  {journal}
  {J. Chem. Phys.}\ }\textbf {\bibinfo {volume} {81}},\ \bibinfo {pages} {511}
  (\bibinfo {year} {1984})}\BibitemShut {NoStop}%
\bibitem [{\citenamefont {Hoover}(1985)}]{hoover1985canonical}%
  \BibitemOpen
  \bibfield  {author} {\bibinfo {author} {\bibfnamefont {W.~G.}\ \bibnamefont
  {Hoover}},\ }\bibfield  {title} {\bibinfo {title} {Canonical dynamics:
  Equilibrium phase-space distributions},\ }\href
  {https://doi.org/10.1103/PhysRevA.31.1695} {\bibfield  {journal} {\bibinfo
  {journal} {Phys. Rev. A}\ }\textbf {\bibinfo {volume} {31}},\ \bibinfo
  {pages} {1695} (\bibinfo {year} {1985})}\BibitemShut {NoStop}%
\bibitem [{\citenamefont {Dai}\ \emph {et~al.}(2010)\citenamefont {Dai},
  \citenamefont {Hou},\ and\ \citenamefont {Yuan}}]{dai2010quantum}%
  \BibitemOpen
  \bibfield  {author} {\bibinfo {author} {\bibfnamefont {J.}~\bibnamefont
  {Dai}}, \bibinfo {author} {\bibfnamefont {Y.}~\bibnamefont {Hou}},\ and\
  \bibinfo {author} {\bibfnamefont {J.}~\bibnamefont {Yuan}},\ }\bibfield
  {title} {\bibinfo {title} {Quantum langevin molecular dynamic determination
  of the solar-interior equation of state},\ }\href
  {https://doi.org/10.1088/0004-637X/721/2/1158} {\bibfield  {journal}
  {\bibinfo  {journal} {The Astrophysical Journal}\ }\textbf {\bibinfo {volume}
  {721}},\ \bibinfo {pages} {1158} (\bibinfo {year} {2010})}\BibitemShut
  {NoStop}%
\bibitem [{\citenamefont {McMahon}\ \emph {et~al.}(2012)\citenamefont
  {McMahon}, \citenamefont {Morales}, \citenamefont {Pierleoni},\ and\
  \citenamefont {Ceperley}}]{mcmahon2012the}%
  \BibitemOpen
  \bibfield  {author} {\bibinfo {author} {\bibfnamefont {J.~M.}\ \bibnamefont
  {McMahon}}, \bibinfo {author} {\bibfnamefont {M.~A.}\ \bibnamefont
  {Morales}}, \bibinfo {author} {\bibfnamefont {C.}~\bibnamefont {Pierleoni}},\
  and\ \bibinfo {author} {\bibfnamefont {D.~M.}\ \bibnamefont {Ceperley}},\
  }\bibfield  {title} {\bibinfo {title} {The properties of hydrogen and helium
  under extreme conditions},\ }\href
  {https://doi.org/10.1103/RevModPhys.84.1607} {\bibfield  {journal} {\bibinfo
  {journal} {Rev. Mod. Phys.}\ }\textbf {\bibinfo {volume} {84}},\ \bibinfo
  {pages} {1607} (\bibinfo {year} {2012})}\BibitemShut {NoStop}%
\bibitem [{\citenamefont {Wang}\ \emph {et~al.}(2013)\citenamefont {Wang},
  \citenamefont {He},\ and\ \citenamefont {Zhang}}]{wang2013thermophysical}%
  \BibitemOpen
  \bibfield  {author} {\bibinfo {author} {\bibfnamefont {C.}~\bibnamefont
  {Wang}}, \bibinfo {author} {\bibfnamefont {X.-T.}\ \bibnamefont {He}},\ and\
  \bibinfo {author} {\bibfnamefont {P.}~\bibnamefont {Zhang}},\ }\bibfield
  {title} {\bibinfo {title} {Thermophysical properties of hydrogen-helium
  mixtures: Re-examination of the mixing rules via quantum molecular dynamics
  simulations},\ }\href {https://doi.org/10.1103/PhysRevE.88.033106} {\bibfield
   {journal} {\bibinfo  {journal} {Phys. Rev. E}\ }\textbf {\bibinfo {volume}
  {88}},\ \bibinfo {pages} {033106} (\bibinfo {year} {2013})}\BibitemShut
  {NoStop}%
\bibitem [{\citenamefont {Stevenson}(1975)}]{stevenson1975thermodynamics}%
  \BibitemOpen
  \bibfield  {author} {\bibinfo {author} {\bibfnamefont {D.~J.}\ \bibnamefont
  {Stevenson}},\ }\bibfield  {title} {\bibinfo {title} {Thermodynamics and
  phase separation of dense fully ionized hydrogen-helium fluid mixtures},\
  }\href {https://doi.org/10.1103/PhysRevB.12.3999} {\bibfield  {journal}
  {\bibinfo  {journal} {Phys. Rev. B}\ }\textbf {\bibinfo {volume} {12}},\
  \bibinfo {pages} {3999} (\bibinfo {year} {1975})}\BibitemShut {NoStop}%
\bibitem [{\citenamefont {Klepeis}\ \emph {et~al.}(1991)\citenamefont
  {Klepeis}, \citenamefont {Schafer}, \citenamefont {Barbee},\ and\
  \citenamefont {Ross}}]{klepeis1991hydrogen}%
  \BibitemOpen
  \bibfield  {author} {\bibinfo {author} {\bibfnamefont {J.~E.}\ \bibnamefont
  {Klepeis}}, \bibinfo {author} {\bibfnamefont {K.~J.}\ \bibnamefont
  {Schafer}}, \bibinfo {author} {\bibfnamefont {T.~W.}\ \bibnamefont
  {Barbee}},\ and\ \bibinfo {author} {\bibfnamefont {M.}~\bibnamefont {Ross}},\
  }\bibfield  {title} {\bibinfo {title} {Hydrogen-helium mixtures at megabar
  pressures: Implications for jupiter and saturn},\ }\href
  {https://doi.org/10.1126/science.254.5034.986} {\bibfield  {journal}
  {\bibinfo  {journal} {Science}\ }\textbf {\bibinfo {volume} {254}},\ \bibinfo
  {pages} {986} (\bibinfo {year} {1991})}\BibitemShut {NoStop}%
\bibitem [{\citenamefont {Vorberger}\ \emph {et~al.}(2007)\citenamefont
  {Vorberger}, \citenamefont {Tamblyn}, \citenamefont {Militzer},\ and\
  \citenamefont {Bonev}}]{vorberger2007hydroge}%
  \BibitemOpen
  \bibfield  {author} {\bibinfo {author} {\bibfnamefont {J.}~\bibnamefont
  {Vorberger}}, \bibinfo {author} {\bibfnamefont {I.}~\bibnamefont {Tamblyn}},
  \bibinfo {author} {\bibfnamefont {B.}~\bibnamefont {Militzer}},\ and\
  \bibinfo {author} {\bibfnamefont {S.~A.}\ \bibnamefont {Bonev}},\ }\bibfield
  {title} {\bibinfo {title} {Hydrogen-helium mixtures in the interiors of giant
  planets},\ }\href {https://doi.org/10.1103/PhysRevB.75.024206} {\bibfield
  {journal} {\bibinfo  {journal} {Phys. Rev. B}\ }\textbf {\bibinfo {volume}
  {75}},\ \bibinfo {pages} {024206} (\bibinfo {year} {2007})}\BibitemShut
  {NoStop}%
\bibitem [{\citenamefont {Militzer}(2013)}]{militzer2013}%
  \BibitemOpen
  \bibfield  {author} {\bibinfo {author} {\bibfnamefont {B.}~\bibnamefont
  {Militzer}},\ }\bibfield  {title} {\bibinfo {title} {Equation of state
  calculations of hydrogen-helium mixtures in solar and extrasolar giant
  planets},\ }\href {https://doi.org/10.1103/PhysRevB.87.014202} {\bibfield
  {journal} {\bibinfo  {journal} {Phys. Rev. B}\ }\textbf {\bibinfo {volume}
  {87}},\ \bibinfo {pages} {014202} (\bibinfo {year} {2013})}\BibitemShut
  {NoStop}%
\bibitem [{\citenamefont {Pfaffenzeller}\ \emph {et~al.}(1995)\citenamefont
  {Pfaffenzeller}, \citenamefont {Hohl},\ and\ \citenamefont
  {Ballone}}]{pfaffenzeller1995miscibility}%
  \BibitemOpen
  \bibfield  {author} {\bibinfo {author} {\bibfnamefont {O.}~\bibnamefont
  {Pfaffenzeller}}, \bibinfo {author} {\bibfnamefont {D.}~\bibnamefont
  {Hohl}},\ and\ \bibinfo {author} {\bibfnamefont {P.}~\bibnamefont
  {Ballone}},\ }\bibfield  {title} {\bibinfo {title} {Miscibility of hydrogen
  and helium under astrophysical conditions},\ }\href
  {https://doi.org/10.1103/PhysRevLett.74.2599} {\bibfield  {journal} {\bibinfo
   {journal} {Phys. Rev. Lett.}\ }\textbf {\bibinfo {volume} {74}},\ \bibinfo
  {pages} {2599} (\bibinfo {year} {1995})}\BibitemShut {NoStop}%
\bibitem [{\citenamefont {Lorenzen}\ \emph {et~al.}(2009)\citenamefont
  {Lorenzen}, \citenamefont {Holst},\ and\ \citenamefont
  {Redmer}}]{lorenzen2009demixing}%
  \BibitemOpen
  \bibfield  {author} {\bibinfo {author} {\bibfnamefont {W.}~\bibnamefont
  {Lorenzen}}, \bibinfo {author} {\bibfnamefont {B.}~\bibnamefont {Holst}},\
  and\ \bibinfo {author} {\bibfnamefont {R.}~\bibnamefont {Redmer}},\
  }\bibfield  {title} {\bibinfo {title} {Demixing of hydrogen and helium at
  megabar pressures},\ }\href {https://doi.org/10.1103/PhysRevLett.102.115701}
  {\bibfield  {journal} {\bibinfo  {journal} {Phys. Rev. Lett.}\ }\textbf
  {\bibinfo {volume} {102}},\ \bibinfo {pages} {115701} (\bibinfo {year}
  {2009})}\BibitemShut {NoStop}%
\bibitem [{\citenamefont {Li}\ \emph {et~al.}(2017)\citenamefont {Li},
  \citenamefont {Zhang}, \citenamefont {Fu}, \citenamefont {Dai}, \citenamefont
  {Chen},\ and\ \citenamefont {Chen}}]{li2007benchmarking}%
  \BibitemOpen
  \bibfield  {author} {\bibinfo {author} {\bibfnamefont {Z.-G.}\ \bibnamefont
  {Li}}, \bibinfo {author} {\bibfnamefont {W.}~\bibnamefont {Zhang}}, \bibinfo
  {author} {\bibfnamefont {Z.-J.}\ \bibnamefont {Fu}}, \bibinfo {author}
  {\bibfnamefont {J.-Y.}\ \bibnamefont {Dai}}, \bibinfo {author} {\bibfnamefont
  {Q.-F.}\ \bibnamefont {Chen}},\ and\ \bibinfo {author} {\bibfnamefont
  {X.-R.}\ \bibnamefont {Chen}},\ }\bibfield  {title} {\bibinfo {title}
  {{Benchmarking the diffusion and viscosity of H-He mixtures in warm dense
  matter regime by quantum molecular dynamics simulations}},\ }\href
  {https://doi.org/10.1063/1.4983057} {\bibfield  {journal} {\bibinfo
  {journal} {Phys. Plasmas}\ }\textbf {\bibinfo {volume} {24}},\ \bibinfo
  {pages} {052903} (\bibinfo {year} {2017})}\BibitemShut {NoStop}%
\bibitem [{\citenamefont {Chang}\ \emph {et~al.}(2023)\citenamefont {Chang},
  \citenamefont {Chen}, \citenamefont {Zeng}, \citenamefont {Wang},
  \citenamefont {Chen}, \citenamefont {Tong}, \citenamefont {Yu}, \citenamefont
  {Kang}, \citenamefont {Zhang}, \citenamefont {Guo}, \citenamefont {Hou},
  \citenamefont {Zhao}, \citenamefont {Yao}, \citenamefont {Ma},\ and\
  \citenamefont {Dai}}]{chang2023direct}%
  \BibitemOpen
  \bibfield  {author} {\bibinfo {author} {\bibfnamefont {X.}~\bibnamefont
  {Chang}}, \bibinfo {author} {\bibfnamefont {B.}~\bibnamefont {Chen}},
  \bibinfo {author} {\bibfnamefont {Q.}~\bibnamefont {Zeng}}, \bibinfo {author}
  {\bibfnamefont {H.}~\bibnamefont {Wang}}, \bibinfo {author} {\bibfnamefont
  {K.}~\bibnamefont {Chen}}, \bibinfo {author} {\bibfnamefont {Q.}~\bibnamefont
  {Tong}}, \bibinfo {author} {\bibfnamefont {X.}~\bibnamefont {Yu}}, \bibinfo
  {author} {\bibfnamefont {D.}~\bibnamefont {Kang}}, \bibinfo {author}
  {\bibfnamefont {S.}~\bibnamefont {Zhang}}, \bibinfo {author} {\bibfnamefont
  {F.}~\bibnamefont {Guo}}, \bibinfo {author} {\bibfnamefont {Y.}~\bibnamefont
  {Hou}}, \bibinfo {author} {\bibfnamefont {Z.}~\bibnamefont {Zhao}}, \bibinfo
  {author} {\bibfnamefont {Y.}~\bibnamefont {Yao}}, \bibinfo {author}
  {\bibfnamefont {Y.}~\bibnamefont {Ma}},\ and\ \bibinfo {author}
  {\bibfnamefont {J.}~\bibnamefont {Dai}},\ }\href@noop {} {\bibinfo {title}
  {Direct evidence of helium rain in jupiter and saturn}} (\bibinfo {year}
  {2023}),\ \Eprint {https://arxiv.org/abs/2310.13412} {arXiv:2310.13412
  [physics.comp-ph]} \BibitemShut {NoStop}%
\bibitem [{\citenamefont {Sjostrom}\ and\ \citenamefont
  {Crockett}(2015)}]{sjostrom2015orbital}%
  \BibitemOpen
  \bibfield  {author} {\bibinfo {author} {\bibfnamefont {T.}~\bibnamefont
  {Sjostrom}}\ and\ \bibinfo {author} {\bibfnamefont {S.}~\bibnamefont
  {Crockett}},\ }\bibfield  {title} {\bibinfo {title} {Orbital-free extension
  to kohn-sham density functional theory equation of state calculations:
  Application to silicon dioxide},\ }\href
  {https://doi.org/10.1103/PhysRevB.92.115104} {\bibfield  {journal} {\bibinfo
  {journal} {Phys. Rev. B}\ }\textbf {\bibinfo {volume} {92}},\ \bibinfo
  {pages} {115104} (\bibinfo {year} {2015})}\BibitemShut {NoStop}%
\bibitem [{\citenamefont {Kang}\ \emph
  {et~al.}(2020{\natexlab{b}})\citenamefont {Kang}, \citenamefont {Hou},
  \citenamefont {Zeng},\ and\ \citenamefont {Dai}}]{kang2020unified}%
  \BibitemOpen
  \bibfield  {author} {\bibinfo {author} {\bibfnamefont {D.}~\bibnamefont
  {Kang}}, \bibinfo {author} {\bibfnamefont {Y.}~\bibnamefont {Hou}}, \bibinfo
  {author} {\bibfnamefont {Q.}~\bibnamefont {Zeng}},\ and\ \bibinfo {author}
  {\bibfnamefont {J.}~\bibnamefont {Dai}},\ }\bibfield  {title} {\bibinfo
  {title} {{Unified first-principles equations of state of deuterium-tritium
  mixtures in the global inertial confinement fusion region}},\ }\href
  {https://doi.org/10.1063/5.0008231} {\bibfield  {journal} {\bibinfo
  {journal} {Matter Radiat. Extremes}\ }\textbf {\bibinfo {volume} {5}},\
  \bibinfo {pages} {055401} (\bibinfo {year} {2020}{\natexlab{b}})}\BibitemShut
  {NoStop}%
\bibitem [{\citenamefont {Kresse}\ and\ \citenamefont
  {Furthm\"uller}(1996)}]{Kresse1996vasp1}%
  \BibitemOpen
  \bibfield  {author} {\bibinfo {author} {\bibfnamefont {G.}~\bibnamefont
  {Kresse}}\ and\ \bibinfo {author} {\bibfnamefont {J.}~\bibnamefont
  {Furthm\"uller}},\ }\bibfield  {title} {\bibinfo {title} {Efficient iterative
  schemes for ab initio total-energy calculations using a plane-wave basis
  set},\ }\href {https://doi.org/10.1103/PhysRevB.54.11169} {\bibfield
  {journal} {\bibinfo  {journal} {Phys. Rev. B}\ }\textbf {\bibinfo {volume}
  {54}},\ \bibinfo {pages} {11169} (\bibinfo {year} {1996})}\BibitemShut
  {NoStop}%
\bibitem [{\citenamefont {Kresse}\ and\ \citenamefont
  {Furthmüller}(1996)}]{Kresse1996vasp2}%
  \BibitemOpen
  \bibfield  {author} {\bibinfo {author} {\bibfnamefont {G.}~\bibnamefont
  {Kresse}}\ and\ \bibinfo {author} {\bibfnamefont {J.}~\bibnamefont
  {Furthmüller}},\ }\bibfield  {title} {\bibinfo {title} {Efficiency of
  ab-initio total energy calculations for metals and semiconductors using a
  plane-wave basis set},\ }\href
  {https://doi.org/https://doi.org/10.1016/0927-0256(96)00008-0} {\bibfield
  {journal} {\bibinfo  {journal} {Comput. Mater. Sci.}\ }\textbf {\bibinfo
  {volume} {6}},\ \bibinfo {pages} {15} (\bibinfo {year} {1996})}\BibitemShut
  {NoStop}%
\bibitem [{\citenamefont {Bl\"ochl}(1994)}]{PAW1994}%
  \BibitemOpen
  \bibfield  {author} {\bibinfo {author} {\bibfnamefont {P.~E.}\ \bibnamefont
  {Bl\"ochl}},\ }\bibfield  {title} {\bibinfo {title} {Projector augmented-wave
  method},\ }\href {https://doi.org/10.1103/PhysRevB.50.17953} {\bibfield
  {journal} {\bibinfo  {journal} {Phys. Rev. B}\ }\textbf {\bibinfo {volume}
  {50}},\ \bibinfo {pages} {17953} (\bibinfo {year} {1994})}\BibitemShut
  {NoStop}%
\end{thebibliography}%

\end{document}